\documentclass[a4paper,fleqn]{cas-dc}

\usepackage{cite}
\usepackage[numbers]{natbib}
\usepackage{subfigure}
\usepackage{amsmath,amssymb,amsfonts}
\usepackage{algorithmic}
\usepackage{graphicx}
\usepackage{afterpage}
\usepackage{textcomp}
\usepackage{xcolor}
\usepackage{mathtools}
\usepackage[many]{tcolorbox}
\usepackage{booktabs}
\usepackage{csvsimple}
\usepackage{longtable}
\usepackage{filecontents}
\usepackage{tabularx}
\usepackage{datatool}
\usepackage{adjustbox}
\usepackage{array}
\usepackage{xtab,booktabs}
\usepackage{tikz}
\usetikzlibrary{fit,arrows,calc,positioning}
\usepackage{color}
\usepackage{colortbl}
\definecolor{mycolor3}{cmyk}{0, 0.708, 0.49, 0.112}
\definecolor{aliceblue}{rgb}{0.64, 0.76, 0.68}
\def\tsc#1{\csdef{#1}{\textsc{\lowercase{#1}}\xspace}}
\tsc{WGM}
\tsc{QE}
\tsc{EP}
\tsc{PMS}
\tsc{BEC}
\tsc{DE}

\colorlet{mygreen}{green!80!black}

    \newcommand{\new}[1]{{\color{black}{#1}}}
    \newcommand{\newer}[1]{{\color{black}{#1}}}
\def\BibTeX{{\rm B\kern-.05em{\sc i\kern-.025em b}\kern-.08em
    T\kern-.1667em\lower.7ex\hbox{E}\kern-.125emX}}
\begin{document}

\let\WriteBookmarks\relax
\def\floatpagepagefraction{1}
\def\textpagefraction{.001}
\shorttitle{E-SC4R: Explaining Software Clustering for Remodularisation}
\shortauthors{AJJ Tan et~al.}

\title [mode = title]{E-SC4R: Explaining Software Clustering for Remodularisation}                      

\author[1]{Alvin Jian Jia Tan}

\ead{alvin.tan@monash.edu}

\address[1]{School of Information Technology, Monash University Malaysia, Jalan Lagoon Selatan, Bandar Sunway, 47500 Subang Jaya, Selangor, Malaysia}

\author[1]{Chun Yong Chong}
\cormark[1]
\ead{chong.chunyong@monash.edu}
\author[2]{Aldeida Aleti}

\ead{aldeida.aleti@monash.edu}

\address[2]{Faculty of Information Technology, Monash University, Clayton 3168, VIC, Australia}

\cortext[cor1]{Corresponding author}



\begin{abstract}
Maintenance of existing software requires a large amount of time for comprehending the source code. The architecture of a software, however, may not be clear to maintainers if up-to-date documentations are not available. Software clustering is often used as a remodularisation and architecture recovery technique to help recover a semantic representation of the software design. \new{Due to the diverse domains, structure, and behaviour of software systems, the suitability of different clustering algorithms for different software systems are not investigated thoroughly.} Research that introduce new clustering techniques usually validate their approaches on a specific domain, which might limit its generalisability. If the chosen test subjects could only represent a narrow perspective of the whole picture, researchers might risk not being able to address the external validity of their findings. This work aims to fill this gap by introducing a new approach, Explaining Software Clustering for Remodularisation (E-SC4R), to evaluate the effectiveness of different software clustering approaches. This work focuses on hierarchical clustering and Bunch clustering algorithms and provides information about their suitability according to the features of the software, which as a consequence, enables the selection of the \newer{most suitable algorithm and configuration that can achieve the best MoJoFM value} from our existing pool of choices for a particular software system. The E-SC4R framework is tested on 30 open-source software systems with varying sizes and domains, and demonstrates that it can characterise both the strengths and weaknesses of the analysed software clustering algorithms using software features extracted from the code. The proposed approach also provides a better understanding of the algorithms’ behaviour by showing a 2D representation of the effectiveness of clustering techniques on the feature space generated through the application of dimensionality reduction techniques. 
\end{abstract}

\maketitle

\section{Introduction}\label{sec:introduction}
Software clustering is one of the software remodularisation and software architecture recovery techniques that has received a substantial amount of attention in recent years \cite{hall2018effectively, RN66, RN67, wu2005comparison, RN69, RN70, andritsos2005information}. \new{There are many goals for software remodularisation, including, but not limited to, representing the high-level architectural view of the analysed software, removing potential technical debt due to suboptimal software structure, and reverse documentation of poorly documented systems.} Software remodularisation can help software developers and maintainers better understand the interrelationships between software components.

As discussed in the work by Teymourian et al. \cite{teymourian2020fast}, most of the software clustering algorithms fall into two main categories which are agglomerative hierarchical \cite{RN66, RN67, chong2013efficient, aghdasifam2020new} and search-based
algorithms \cite{mitchell2002heuristic, mitchell2006,prajapati2020harmony}. In general, software clustering works by choosing from a collection of software entities (methods, classes, or packages) and then forming multiple groups of entities such that the entities within the same group are similar to each other while being dissimilar from entities in other groups. By dividing and grouping software entities based on their functionality, these groups or clusters can be recognised as functionally similar subsystems, which can be used to represent the software architecture of the system. Ultimately, the high-level architecture view of the software system will aid software maintainers in implementing new functionalities or make changes to existing code through better comprehension of the software design.

Due to their capabilities to aid in architecture recovery and software remodularisation, software clustering techniques have been widely investigated and a large number of techniques have been introduced \cite{maqbool2007hierarchical, hall2018effectively}. These techniques differ greatly in terms of the chosen common clustering features, similarity measures, clustering algorithm, and evaluation metric \cite{maqbool2007hierarchical,shtern2010comparability,mitchell2006}. \new{There is a vast variety of software systems from different domains with unique structures, characteristics, and behaviour. However, the suitability of different clustering algorithms for different software systems are not investigated thoroughly.}

Different clustering algorithms tend to produce semantically different clustering results. For instance, if classes are chosen as the basis to perform software clustering, the clustering feature extraction method will only look at class-level interaction between those classes. Subsequently, the clustering results produced by the class-level clustering algorithm will be completely different from a method-level clustering algorithm, although both results might be equally feasible. Furthermore, comparing software clustering algorithms within the same level of granularity is also not straightforward, due to different fitness functions and cluster validity metrics employed by different algorithms \cite{chong2017automatic,chong2013efficient}. Even if we were to compare the effectiveness of the clustering algorithms from the same family (i.e., agglomerative hierarchical clustering), there are still different ways to configure them (i.e. different distance metrics, different linkage algorithms, and different validity indices for hierarchical clustering algorithm). It is then up to the researchers to choose a software clustering evaluation method depending upon if they are able to produce the reference decomposition or model, to evaluate the effectiveness of the clustering results. Almost all of the existing studies in software clustering only emphasised on the advantages and benefits brought upon by the proposed clustering technique, while limited studies suggest the limitation of their approaches \cite{shtern2012,shtern2010comparability,maqbool2007hierarchical}. Therefore, it raises a question as to how software clustering algorithms are evaluated. 

Most studies which introduce new clustering algorithms often only evaluate their approach on a specific set of problem instances \cite{maqbool2007hierarchical,chong2017automatic,shtern2012}. Different from existing studies, this work aims to provide a better understanding of which software/code features \newer{(i.e., lines of code, number of methods, coupling between objects, depth inheritance)} are related to the performance of clustering algorithms, and whether the software/code features can be used to select the \newer{most suitable} clustering algorithm. Our work is inspired from similar research efforts in optimisation and search-based software testing~\cite{Oliveira2019,oliveira2018mapping}. 

\begin{figure*}[!ht]
    \centering
    \includegraphics[width=1\linewidth]{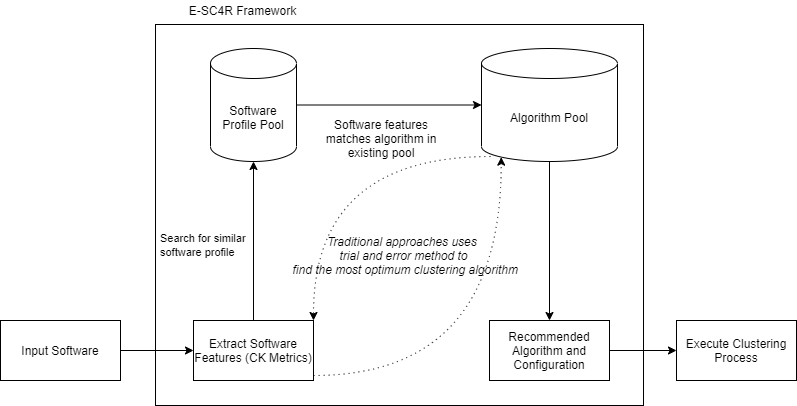}
    \caption{\new{E-SC4R framework design and workflow.}}
    \label{fig:e-sc4r_project_design_tool}
\end{figure*}

\new{
This work aims to fill the gap by introducing a new approach that evaluates the effectiveness of \newer{software clustering algorithms by providing information on their strengths and weaknesses} according to the software or code features. \newer{This can be used for profiling of selected clustering algorithms}, enabling the selection of the most suitable algorithm and configuration \newer{that can achieve the best MoJoFM value} from our existing pool of choices for a particular profile of a software system. \newer{The proposed framework also provides a clear understanding of the algorithm's behaviour by showing a 2D representation of the effectiveness of software clustering techniques on the feature space through the application of dimensionality reduction techniques. This can be extended to algorithm improvements, considering one of the aims of the proposed approach is to reveal the weaknesses of clustering algorithms.} Using the proposed framework, the pool of chosen software clustering algorithms only requires profiling to be done for the first time, in order for the proposed framework to recommend \newer{a suitable} algorithm and configuration from our existing pool of choices . Software systems that exhibit characteristics that match a profile from the pool of choices (software clustering algorithm) will be recommended with the respective clustering algorithm to improve the overall efficiency and effectiveness. The entire workflow is summarised in Figure~\ref{fig:e-sc4r_project_design_tool}. Different from traditional software clustering research that uses a trial and error approach to identify \newer{a suitable} clustering algorithm and its configuration (dashed line in Figure~\ref{fig:e-sc4r_project_design_tool}), the proposed framework only requires the developer or researcher to extract the software features of the software to be remodularised, in order for the proposed framework to recommend the most suitable clustering algorithm. 
}

Note that in this paper, we only focus on comparing and evaluating the effectiveness of different variants of agglomerative hierarchical clustering algorithms and Bunch clustering algorithm because i.) agglomerative clustering and Bunch (which is a search-based software clustering algorithm) are two of the most popular clustering algorithms as discussed in \cite{teymourian2020fast}, and ii.) results produce by different family of software clustering algorithms exhibit significantly different structure and behaviour, which are difficult to compare directly. \newer{To avoid confusion, the term \textit{clustering features} used in this paper refers to the \new{features extracted from the chosen clustering entities (classes)}, while \textit{software features} refers to the characteristic of the software/code such as the lines of code, number of public methods, number of static methods \new{and coupling between objects}.}

In essence, the proposed approach can be used to characterise the software/code features that have an impact on the effectiveness of clustering algorithms. It is a known fact that the selection of clustering entities and clustering features will directly influence the final clustering results. If we could understand and correlate the relationships between software/code features and the effectiveness of software clustering algorithms, it is then possible to choose the most \newer{suitable} clustering algorithm and configuration based on the profiled software/code features. We show how such software/code features can be measured, and how the footprints of software clustering algorithms (regions where clustering algorithms' strengths are expected) can be visualised across the inspected software components. 

The research questions are: 
       \begin{itemize}
       \item  [RQ1] How can we identify the strengths and weaknesses of clustering \newer{algorithms} for the remodularisation of software systems?
        \item [RQ2] How can we select the most \newer{suitable} clustering technique from a portfolio of hierarchical and Bunch clustering \newer{algorithms}?
       \end{itemize}
       
\section{The E-SC4R Framework}

The E-SC4R framework provides a way for objective assessment of the overall effectiveness of software clustering techniques. Understanding the effectiveness of a software clustering technique is critical in selecting the \newer{most suitable} technique for a particular software, avoiding trial and error application of software clustering techniques.

The purpose of E-SC4R is the ability to \textbf{identify the most \newer{suitable} algorithm and configuration from our existing pool of choices for software clustering.}
 
The approach involves two main parts: \textbf{Strengths and Weaknesses of Clustering Techniques}, which learns significant software features, such as lines of code, cohesion, coupling, and complexity, that reveal why certain remodularisation problems are hard. The first part visualises footprints of clustering techniques which expose their strengths and weaknesses, and \textbf{Clustering Technique Selection}, which addresses the problem of selecting the \newer{most suitable} technique for software remodularisation. The approach we propose has two main goals:
\begin{itemize}
    \item to help designers of software clustering techniques gain insight into why some techniques might be more or less suited to remodularise certain software systems, thus devising new and better techniques that address any challenging areas, and
    \item to help software developers select the most effective clustering technique for their software systems.
\end{itemize}

\tikzstyle{block} = [rectangle, draw, fill=mycolor3!15,text width=14em,  text centered, minimum height=4em, thick, rounded corners=4pt]

\tikzstyle{block1} = [rectangle, draw, fill=white,text width=14em,  text centered, minimum height=4em, thick, rounded corners=4pt]

\tikzstyle{big} = [rectangle, draw, inner sep=0.5cm]

\tikzstyle{line} = [draw, -latex',thick]

\begin{figure*}[hbtp]
 \centering
 \begin{tikzpicture}
  [node distance=3cm and 2.5cm]
\node [block1](1) {\textbf{$s\in S$} \\ Software Systems};
\node [block, right=of 1] (2) {\textbf{$Q(F(T)) \in R$ \\ } Strengths and Weaknesses of Clustering Techniques};

\node [block1, below=1cm of 1] (3) {$C(s) \in Y$  \\ Software Clustering Techniques};
\node [block1, right=of 3] (4) {\textbf{$I(C(s)) \in R$ \\} Clustering Technique Performance Indicator};
\node [block, right=of 2, xshift=-1cm] (5) {\textbf{$A(Q(F(T))) \in T$ \\} Clustering Technique Selection};

\draw [thick, ->] (1) -- (2) node[midway, above] {software features};
\draw [thick, ->] (1) -- (3) node[midway, right] {source code, dissimilarity matrix};
\draw [thick, ->] (3) -- (4) node[midway, above] {clustering results};
\draw [thick, ->] (4) -- (2) node[midway, right] {MoJoFM results};
\draw [thick, ->] (2) -- (5) node[midway, above] {footprints};

\end{tikzpicture}
\caption{\new{\normalsize An overview of E-SC4R.\label{fig:method}}}
 \end{figure*}
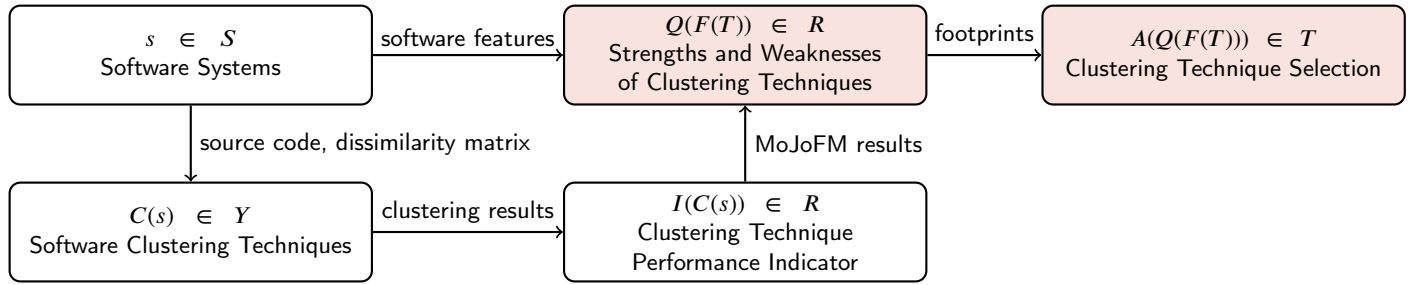
 
An overview of the E-SC4R framework is presented in Figure~\ref{fig:method}. The boxes represent the artefacts, while the arrows are the processes/steps for creating the artefacts. In the following subsections, we describe our framework in more detail for each artefact/step.


\subsection{Software Systems}

The software systems, defined as \textbf{$s\in S$} in Figure~\ref{fig:method} are software systems to be remodularised by researchers using software clustering techniques to recover a high-level abstraction view of the software architecture. 

\new{
The software systems used in this study are chosen in a pseudo-random manner, with a mixture of GitHub projects and Apache projects.

The following process is used when selecting the projects from the main GitHub and Apache repository.

\begin{enumerate}
    \item The search parameters are set to filter out Java-based projects at https://github.com/topics/java,
    \item Sort projects by the number of stars,
    \item Projects are manually chosen if they meet the following criteria,
    \begin{itemize}
        \item Have at least 10 commits in the past year,
        \item README.MD, Project Title, "About" and comments are written in English.
    \end{itemize}
\end{enumerate}

\newer{In order to collect a good representation of active and popular open source projects hosted on Github, we sorted the software systems by the number of stars. Another selection criteria to generate a reliable ground truth (discuss in more detail in Section 3.4) is that the selected projects must have a reasonable amount of commits and releases. Selecting active and popular projects based on Github stars allows us to identify such projects. Hence, in this study, we only chose the top 30 projects ranked by the number of stars.}
In order to make sure that the selected project is still currently active, we also made sure that the chosen projects that have at least 10 commits in 2021 (the year in which we conducted the experiment). 
}

\subsection{Software Clustering}
Software clustering techniques are defined as \textbf{$C(s) \in Y$ } in Figure~\ref{fig:method}, where a specific clustering technique $C$, which is a subset of $Y$, applied on software $s$.

Depending on the different interpretation of a meaningful and effective software clustering result, different clustering algorithms proposed in the existing literature address the software remodularisation and architecture recovery problem in a different manner. While these algorithms usually aim to achieve common goals, i.e., improve the software modularity, quality, and software comprehension, they usually produce significantly different clustering results but at the same time, might also offer equally valid high-level abstraction views of the analysed software \cite{factbase}. Hence, it is necessary to discuss the working principle of some of the widely adopted software clustering algorithms available in the current literature.

\subsubsection{Search-based Software Remodularisation Algorithms}
Search-based approaches have been successfully utilised to address the software remodularisation problem. In general, the workflow of search-based clustering algorithms consists of the following steps\new{ \cite{RN67,mitchell2006}}:

\begin{enumerate}
\item Generate a random seed solution based on some initial parameters,
\item Explore the neighbourhood structure of the solution. If a neighbour has a better fitness, it becomes the new solution,
\item Repeat step (2) until the neighbourhood of the current candidate solution offers no further improvement to fitness function (local optimal),
\item Restart step (1) with different seed to find better solution and fitness (global optimal).

\end{enumerate}\par

Numerous existing studies that adopt search-based approaches are based on the work by Mitchell and Mancoridis \cite{mancoridis1998,mitchell2006}, where their approaches have been implemented as part of the Bunch software clustering tool. The authors proposed the notion of module dependency graph (MDG) as the basis of their clustering entities. In their context, module represents source code entities that encapsulate data and functions that operate on the data (e.g., Java/C++ classes, C source code files). Dependencies between the modules, on the other hand, are binary relations between the source code entities, depending upon the programming language used to implement the analysed software systems (e.g., function/procedure invocation, variable access, and inheritance). The Bunch clustering algorithm works by generating a random partition of the MDG which is then re-partition systematically by examining neighbouring structures in order to find a better partition. When an improved partition which yields better intra-cluster cohesion and inter-cluster coupling is found, the process repeats by using the newly found partition as the basis for finding the next improved partition. The algorithm stops when it cannot find a better partition. 

Based on the work by Mitchell and Mancoridis, Harman et al. propose a search-based optimisation method to search for the best optimum partition \cite{harman2005}. In their work, Harman et al. proposed a new objective function for the search-based problem such that for each pair of modules in a partition/cluster, the optimisation score is incremented if there is a dependency between the modules; otherwise, it is decremented. The authors do not consider inter-cluster coupling and only focuses on intra-cluster cohesion. Experiment results show that their proposed approach can tolerate noises better than the Bunch clustering algorithm \cite{mancoridis1998} and reach optimum results faster than Bunch-guided approach. 

In the work by Beck and Diehl \cite{beck2013impact}, the authors discussed that clustering algorithms based on the Bunch tool often only rely on the structural and static information of the source code in order to measure the similarity and dependencies among software entities. The authors attempted to enrich the structural data with some evolutionary aspects (historical data) of the analysed software such as size of packages, ratio of code to comments, and the number of download. Experiment results show that using evolutionary data alone does not perform better than the traditional clustering algorithms that utilise structural data. It is only when both data are integrated, the clustering results achieve much higher accuracy when compared against the reference model. 


\subsubsection{Hierarchical Clustering Algorithms}
On the other hand, hierarchical clustering iteratively merges smaller clusters into larger ones or divide large clusters into smaller ones, depending on whether it is a bottom-up or top-down approach. Merging or dividing operations are usually dependent on the clustering algorithm used in the existing studies. In general, hierarchical clustering algorithms can be divided into two main approaches, divisive (top-down) and agglomerative (bottom-up) hierarchical clustering algorithms. 

Divisive clustering is based on a top-down hierarchical clustering approach where the clustering process starts at the top with all data in one big cluster. \newer{The approach recursively splits a cluster into two sub-clusters, starting from the main dataset.} \new{For problem domains with a fixed number of top levels, using flat algorithms such as $K$-mean yield lower computational complexity because divisive clustering is linear in the number of clusters \cite{schutze2008introduction}. Although the computational complexity of divisive clustering} is lower than agglomerative clustering \newer{under normal circumstances}, complete information about the global distribution of the data is needed when making the top-level clustering decisions \cite{dhillon2002enhanced}. Most of the time, software maintainers are not involved in the earlier software design phases. If the software documentations are not up-to-date, it is hard for maintainers to identify the ideal number of software packages (or the number of clusters in the context of software clustering) before any attempt to remodularise any software systems. 

On the other hand, the work by Wiggerts \cite{wiggerts1997using} discussed how agglomerative clustering, a bottom-up clustering approach would be helpful to software engineers in remodularising legacy and poorly documented software systems. According to the author, the working principle of agglomerative clustering is actually similar to reverse engineering where the abstractions of software design are recovered in a bottom-up manner. Agglomerative hierarchical clustering starts by placing each cluster entity (usually code or classes in the context of software clustering) in a cluster on its own. At each iteration as we move up the hierarchy, two of the most similar clusters from the lower layer are merged and the number of clusters is reduced by one. The decision to merge which clusters differs depending on the similarity measure and the linkage algorithm used, i.e., nearest neighbour between two clusters, furthest neighbour between two clusters, or average distance between two clusters. 

Once the two chosen clusters have been merged, the strength of similarity between the newly formed cluster and the rest of the clusters are updated to reflect the changes. The merging process will continue until there is only one cluster left. The results of agglomerative clustering are usually presented in a tree diagram, called dendrogram. A dendrogram shows the taxonomic relationships of clusters produced by the clustering algorithm. Cutting the dendrogram at a certain height produces a set of disjoint clusters.  

In this paper, we will focus on examining the strength and weakness of different configurations for agglomerative hierarchical clustering algorithm and Bunch algorithm because they are one of the most widely used generic clustering algorithm and search-based algorithms in the existing literature \cite{maqbool2007hierarchical,chong2013efficient,beck2016identifying,beck2013impact,naseem2019euclidean}. It will be interesting to compare the performance of two different families of software remodularisation techniques and examine their suitability on different behaving datasets.

\subsection{Clustering Technique Performance Indicator}

The Clustering Technique Performance Indicator, denoted as \textbf{$I(C(s)) \in R$} takes as input the clustering results generated by a clustering algorithm \textbf{$C(s)$} for a particular software system \textbf{$s\in S$}. There exist various ways to evaluate the performance of software clustering algorithms. Typically, the clustering results generated from a clustering algorithm \textbf{$C(s)$} are measured against a reference model, which refers to a known good clustering result or a reliable reference that can act as a baseline for comparison. Hence, performance indicator typically measures the similarity of the clustering results against the reference model. 

One way of accessing the reference model is through feedback from domain experts in the analysed system, for instance, the original designer, system architect, or senior developers directly involved in the development of the software. However, this approach is difficult to realise because the software maintainers are usually not involved in the initial design and development of the maintained software.

In existing works, when there are no inputs from domain experts to create a reference model or ground truth, several authors have chosen to use the directory structure or package structure of the analysed software to create an artificial ground truth or reference model \cite{wu2005comparison,beck2016identifying,beck2013impact}. This method is less expensive compared to retrieving the reference model from domain experts because it can usually be automated. However, the reliability of using the directory or package structure of the analysed system is strongly dependent on the skills and experience of the software developers because it is assumed that software developers follow the best practices of putting functional relevant and similar classes into the same package directory.

If an artificial ground truth or reference model can be retrieved, we can then compare it with the clustering results generated by a clustering algorithm \textbf{$C(s)$} to measure the extent to which two given decompositions of the software are similar to each other. The work by Alsarhan \cite{Alsarhan2020} discussed that one of the most popular performance indicator for software clustering results is the MoJo family of metrics \cite{wen2003optimal}:

\begin{equation}
\label{eq:mojofm}
    \text{MoJoFM}(A,B) = (1- \dfrac{mno(A,B)}{\max(mno(\forall A, B))}) \times 100\%
\end{equation}

where $mno(A,B)$ is the minimum number of Move or Join operations needed to transform the clustering result $(A)$ into ground truth $(B)$, and $max(mno(\forall A, B)$ is the maximum number of operations to transform the clustering result to ground truth. MoJoFM return $0$ if the clustering result is very different from the ground truth, and return $100$ if the clustering result is identical to the ground truth. We will be using MoJoFM as the Clustering Technique Performance Indicator \textbf{$R$} in this paper.

\subsection{Strengths and Weaknesses of Clustering Techniques}
One of the important steps in E-SC4R is identifying features of software systems \newer{$f(s) \in F$ }that have an impact on the effectiveness of software clustering techniques. Software features such as lines of code, number of public methods \newer{and number of static methods} are problem dependent and must be chosen such that depending on the type of target software systems \textbf{$s \in S$}, any known structural properties of the software systems are captured, and any known advantages and limitations of the different software clustering algorithm are related to the features. 

For each version of the analysed project, software features are extracted using the Java code metrics calculator, which is publicly available online \cite{CKmetric}. The tool is capable of calculating simple size metrics such as the number of methods, lines of code, and number of private fields, to more complex measures such as depth of inheritance, coupling between object, and other CK suite of metrics. In total, there are 40 metrics that we extracted using the tool. These metrics are calculated at the class level. To aggregate the calculated metrics at the project level, we calculate the max, mean, standard deviation, and sum over all classes of a project, resulting in $40 \times 4 = 160$ metrics for each project.

\new{Table~\ref{tab:features} shows some of the metrics used in this paper as well as their definitions. The full list of metrics along with its description can be found on the tool's GitHub page.\footnote{https://github.com/mauricioaniche/ck.} The set of features listed in Table~\ref{tab:features} is only a subset of the total number of features used in the paper, where we list some of the more general and widely used software features.}

As such, the terms software metrics and software features will be used interchangeably in this paper \new{to denote the metrics that represent different features of the analysed software}.

\begin{table*}[h!]
    \caption{Description of software features.}
    \label{tab:features}
\renewcommand{\arraystretch}{1.1}
    \begin{tabular}{l|p{12cm}}
    \hline
\multicolumn{2}{c}{\textbf{Object oriented features}}\\\hline
 WMC& {McCabe's complexity. It counts the number of branch instructions in a class.} \\
 DIT& {Depth Inheritance Tree, counts the number of parent a class has.}\\
 Number of Fields& {Counts the number of fields. Specific numbers for total number of fields, static, public, private, protected, default, final, and synchronised fields.}\\
 Number of Methods& {Counts the number of methods. Specific numbers for total number of methods, static, public, abstract, private, protected, default, final, and synchronised methods. Constructor methods also count here.}\\
 Number of visible methods& {Counts the number of visible methods. A method is visible if it is not private.}\\
 NOSI& Number of static invocations, counts the number of invocations to static methods.\\
 RFC& Response for a class, counts the number of unique method invocations in a class.\\
 LOC& Lines of code. It counts the lines of count, ignoring empty lines and comments (i.e., it's Source Lines of Code, or SLOC).\\
 TCC& Tight Class Cohesion, measures the cohesion of a class with a value range from 0 to 1. TCC measures the cohesion of a class via direct connections between visible methods, two methods or their invocation trees access the same class variable.\\
 LCC& Loose Class Cohesion, similar to TCC but it further includes the number of indirect connections between visible classes for the cohesion calculation. Thus, the constraint LCC >= TCC holds always.\\
 No. of Returns& The number of return instructions.\\
 No. of Loops & The number of loops (i.e., for, while, do while, enhanced for).\\
 No. of Comparisons& The number of comparisons (i.e., == and !=). \\
 No. of try/catches& The number of try/catches.\\
 No. of () expressions& The number of expressions inside parenthesis.\\
 String literals& The number of string literals (e.g., "John Doe"). Repeated strings count as many times as they appear.\\
 Quantity of Number& The number of numbers (i.e., int, long, double, float) literals.\\
 Quantity of Math Operations& The number of math operations (times, divide, remainder, plus, minus, left shit, right shift).\\
 Quantity of Variables& Number of declared variables.\\
 Max nested blocks& The highest number of blocks nested together.\\
 Number of unique words& Number of unique words in the source code. Counts the number of words in a method/class, after removing Java keywords. Names are split based on camel case and underline.\\
 Number of Log Statements& Number of log statements in the source code. The counting uses REGEX compatible with SLF4J and Log4J API calls.\\
 Has Javadoc& Boolean indicating whether a method has javadoc.\\
 Usage of each variable& How often each variable was used inside each method.\\
 Usage of each field& How often each local field was used inside each method, local field are fields within a class (subclasses are not included). Also indirect local field usages are detected, indirect local field usages include all usages of fields within the local invocation tree of a class e.g. A invokes B and B uses field a, then a is indirectly used by A.\\
 Method invocations& All directly invoked methods, variations are local invocations and indirect local invocations.\\

 \hline
 \end{tabular}
 \end{table*}

Using the results gathered from clustering algorithms and software feature extraction, we can create the footprint visualisations of each clustering algorithm in order to identify the most significant software features that have an impact on its effectiveness.

E-SC4R identifies software features that have an impact on the effectiveness of software clustering techniques. The clustering results can be affected by the program structure and/or source code, the complexity of dependencies between classes, information about the input/output data space, and information dynamically obtained from program execution. All these aspects may influence the suitability of software clustering techniques for a particular software system. 

In E-SC4R, a subset of features is considered significant if they result in an instance space -- as defined by the 2-dimensional projection of the subset of features -- with software systems where a particular clustering technique performs well being clustered together. The software instances are initially projected in the 2D instance space in such a way that if two software systems are similar according to some features, they are closer together, and if they are dissimilar, then they are far apart. Since we focus on arranging the software systems in a space where the instances of a technique is effective are separated from the ineffective ones, we represent a software system as a vector of the most significant features that are likely to correlate with a clustering technique's effectiveness. E-SC4R identifies software features that are able to create a clear separation in instance space, such that we can clearly see the different clusters of software systems where the techniques are effective. We refer to these clusters as \textit{clustering technique footprints}.

\newer{
The most significant features are determined in 2 steps \cite{aleti2020apr}. Firstly, MoJoFM is determined as the performance metric to measure the effectiveness of the clustering algorithm. After identifying the most suitable clustering algorithm, the next step is to identify software features \new{(e.g. lines of code, number of methods, coupling between objects and depth inheritance)} that influence the suitability of the clustering algorithm (i.e. Why a certain clustering algorithm is the most suitable for a specific software system? Does having a high value for lines of code and number of methods in the analysed project contribute to a certain clustering algorithm's suitability?). Next, the genetic algorithm will try to search for the feature sets that are the best in predicting the most suitable algorithm, as given below:
\begin{enumerate}
    \item Select a set of software features,
    \item Generate an instance space using PCA for dimension reduction, as described below,
    \item Fitness of the set of software features is evaluated
    \item If the set of software features is not suitable, return to Step 1.
\end{enumerate}
Section \ref{ga_process_subsection} contains a detailed example on how this process is carried out.}

\newer{The genetic algorithm will search the space for possible subsets and determine the optimal subset with the classification accuracy on an out-of-sample test set used as the fitness function.} The instance space is generated in iterations, until an optimal subset of features is found \cite{munoz2018instance}. A subset of the features is considered of high quality if they result in a footprint visualisation with distinct clusters. The best subset of features is the one that can best discriminate between high performing and low performing clustering algorithms. 

After the most significant software features (lines of code, number of methods, etc.) are identified, these features are then used as an input into SVM to capture the relationship between the selected features and the \newer{clustering algorithms}.

Similar to our previous work~\cite{oliveira2018mapping,Oliveira2019}, we use principal component analysis (PCA) as a method for projecting the software instances to two dimensions, while making sure that we retain as much information as possible. PCA rotates the data to a new coordinate system $\mathbf{R}^k$, with axes defined by linear combinations of the selected features. The new axes are the eigenvectors of the covariance matrix. The subset of features that have large coefficients and therefore contribute significantly to the variance of each Principal Component (PC), are identified as the significant features. Retaining the two main PCs, the software instances are then projected on this 2D space. Following a similar approach to previous work on dimensionality reduction~\cite{smith2014towards}, we accept the new two dimensional instance space as adequate if most of the variance in the data is explained by the two principal axes. The two principal components are then used to visualise the footprints of the clustering technique. The footprint indicates the area of strength (software instances on which the clustering technique is effective) of each clustering technique.  This step provides an answer to the first research question, \textit{RQ1 : How can we identify the strengths and weaknesses of clustering techniques for the remodularisation of software systems?}

\subsection{Clustering Technique Selection}

Finally, E-SC4R is used to predict, based on the most significant software features, the most effective clustering technique for new remodularisation and architecture recovery problems. This step answers the second research question: \textit{RQ2: How can we select the \newer{most suitable} clustering technique from a portfolio of hierarchical and Bunch clustering algorithms?} This is achieved by modelling the relationship between software features and clustering technique effectiveness by employing a Support Vector Machine~\cite{gholami2017support} to learn this relationship. E-SC4R uses the two-dimensional space as an input to the Support Vector Machine to learn the relationship between the software features and remodularisation method performance. 

The SVM is C-classification. The cost C in the regularisation term and the RBF hyper-parameter $\gamma$ are tuned via grid search in [1,10] and [0,1] respectively.

\new{We use 10-fold cross validation to train the model and assess the model generalisation ability. The cross-validated Mean Squared Error and Error Rate are used as estimates of the model generalisation ability in classification.} At the end of this process, E-SC4R creates a model that can select the most effective technique for remodularisation based on the features of software programs. This model can be retrained and extended further with new remodularisation techniques and software features.

\section{Experimental Design}

In this chapter, we discuss the way how the experiment is designed to examine the strength and weakness of different configurations for agglomerative hierarchical clustering algorithm and Bunch algorithm.

\subsection{Agglomerative Hierarchical Software Clustering}
\begin{figure*}[!ht]
    \centering
    \includegraphics[width=0.75\linewidth]{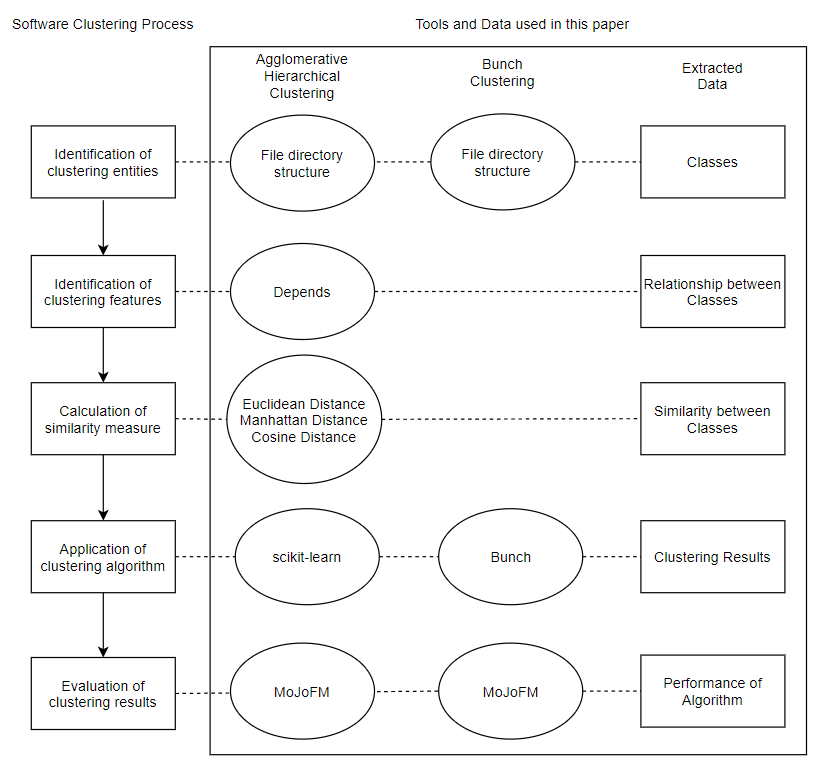}
    \caption{\new{Agglomerative hierarchical and Bunch clustering processes.}}
    \label{fig:agglo_bunch_process}
\end{figure*}

The process of agglomerative hierarchical clustering can be summarised in the following steps, \new{illustrated in Figure \ref{fig:agglo_bunch_process}}.

\begin{enumerate}
    \item Identification of clustering entities,
    \item Identification of clustering features,
    \item Calculation of similarity measure,
    \item Application of clustering algorithm,
    \item Evaluation of clustering results.
\end{enumerate}

\paragraph{Identification of clustering entities}
In software clustering, the typical choices of entities are in the form of methods, classes, or packages because they represent the basic components and functionalities of a software system. In this paper, we focus on representing the software at the class-level because in object-oriented software systems, classes are the main building blocks that contain the implementation details of the examined components. 

\paragraph{Identification of clustering features}
The similarities between entities are determined based on their characteristics or clustering features extracted from the available information. Extracting dependencies between code entities is critical in architecture recovery \new{because it helps to understand the static and dynamic relationships between them}. Several existing studies have proposed different methods to extract dependencies from software entities at different levels of granularity, including code, class, and package levels \cite{jin2019enre,patel2009software,lutellier2015comparing}. The work by Jin et al., \cite{jin2019enre} in particular, released their open-source dependency extraction tool, \textit{Depends} which is capable of gathering syntactical relations among source code entities such as files and methods.\new{\footnote{https://github.com/multilang-depends/depends}} In this work, we choose to utilise \textit{Depends} to extract dependencies between classes in the analysed software in order to improve the replicability of our research findings. Depends extracts the following dependency types:
\new{
\begin{enumerate}
    \item Import,
    \item Contain,
    \item Parameter,
    \item Call,
    \item Return,
    \item Throw,
    \item Implement,
    \item Extend,
    \item Create,
    \item Use,
    \item Cast,
    \item ImplLink,
    \item Annotation,
    \item Mixin.
\end{enumerate}

\new{We are then able to generate \new{an} $NxN$ ($N$ = number of classes) matrix that denotes the relationships between all the classes.
Sample data that \newer{is} extracted from \emph{Depends} are shown in Table~\ref{tab:depends_extracted_data}, which are then aggregated into relationships between clustering entities shown in Table~\ref{tab:depends_matrix}.
}}

\begin{table*}
    \centering
    \caption{\new{Examples of relationships extracted from Depends.}}
    \begin{tabular}{|c|c|c|c|c|c|c|c|c|c|}
    \hline
    src & dest & Cast & Call & Return & Use & Contain & Import & Extend & Implement \\
    \hline
    a.java & b.java & 1 & 8 & 4 & 9 & 0 & 0 & 6 & 3\\
    \hline
    b.java & c.java & 1 & 0 & 0 & 8 & 1 & 0 & 1 & 5\\
    \hline
    \end{tabular}
    \label{tab:depends_extracted_data}
\end{table*}

\begin{table*}
    \centering
    \caption{\new{Examples of relationships between clustering entities aggregated from Table~\ref{tab:depends_extracted_data}.}}
    \begin{tabular}{|c|c|c|c|c|}
    \hline
     & a.java & b.java & c.java & d.java\\
    \hline
    a.java & 0 & 31 & 11 & 9 \\
    \hline
    b.java & 31 & 0 & 16 & 4 \\
    \hline
    c.java & 11 & 16 & 0 & 3 \\
    \hline
    d.java & 9 & 4 & 3 & 0 \\
    \hline
    \end{tabular}
    \label{tab:depends_matrix}
\end{table*}

\paragraph{Calculation of similarity measure}
The next step is to ascertain the similarity between entities by referring to the clustering features identified in the previous step. In this paper, we choose to use distance measures because we are able to quantify the strength of dependencies between classes with the aid of \textit{Depends}. We do not attempt to distinguish between the different type of dependencies identified by \textit{Depends}, but instead aggregate all the dependencies to represent an aggregated strength of dependency between two classes. In order to generate the distance matrix, the following distance measures were taken into consideration to compute the dissimilarity between each class in the examined software:

\begin{itemize}
    \item Euclidean distance : least squares, minimising the sum of the square of the differences between a pair of classes
    
    \begin{equation}
        d(x,y) = {\sqrt{\sum_{i=1}^{n} (x_i - y_i)^2}}
    \end{equation}
    
    where $n$ = number of classes, $x_i$ and $y_i$ are the classes of vectors $x$ and $y$ respectively in the two-dimensional vector space.
    
    \item Manhattan distance : least absolute deviations, minimising the sum of the absolute differences between a pair of classes
    
    \begin{equation}
        d(x,y) = {\sum_{i=1}^{n}}  |x_i - y_i|
    \end{equation}

    \item Cosine distance : the cosine of the angle between a pair of classes
    
    \begin{equation}
        d(x,y) = 1 - {\frac{\sum_{i=1}^{n} x_i y_i}{\sqrt{\sum_{i=1}^{n} x_i^2}{\sqrt{\sum_{i=1}^{n} y_i^2}}}}
    \end{equation}
    
\end{itemize}

\new{These distance/similarity measures are chosen because they have been proven to be effective in measuring the similarity between software components in some of the related studies \cite{maqbool2007hierarchical} \cite{tsantalis_chatzigeorgiou_2009}.}

\paragraph{Application of clustering algorithm}

A clustering algorithm is needed to decide upon how and when to merge two clusters. Depending on the algorithm used, certain algorithms merge the most similar pair first while others merge the most dissimilar first. Once the two chosen clusters have been merged, the strength of similarity or dissimilarity between the newly formed cluster and the rest of the clusters are updated to reflect the changes. It is very common that during hierarchical clustering, there exist more than two entities which are equally similar or dissimilar. In this kind of scenario, the selection of candidate entities to be clustered is arbitrary \cite{maqbool2007hierarchical}. 

In this work, we use the following three linkage algorithms \cite{maqbool2007hierarchical}:

\begin{itemize}
    \item Single Linkage Algorithm - defines the similarity of two chosen clusters as the maximum similarity strength among all pairs of entities (classes) in the two clusters
    \item Average Linkage Algorithm - defines the similarity measure between two clusters as the arithmetic average of similarity strengths among all pairs of entities (classes) in the two clusters
    \item Complete Linkage Algorithm - defines the similarity of two chosen clusters as the minimum similarity strength among all pairs of entities (classes) in the two clusters
\end{itemize}

There exists many other newer algorithms that are proposed for software architecture recovery. \new{Currently, we are only including these three basic linkage algorithms to demonstrate E-SC4R's ability to identify the \new{\newer{most suitable} algorithm and configuration from our existing pool of choices}. Newer algorithms would be added to E-SC4R in future iterations.}

Apart from that, we also attempt to determine the optimum range of number of clusters for each of the chosen hierarchical clustering algorithms. In hierarchical clustering, the final output is represented in a dendrogram, which is a tree diagram that shows the taxonomic relationships of clusters of software entities produced by hierarchical clustering. The distance at which the dendrogram tree is cut determines the number of clusters formed. Cutting the dendrogram tree at a higher distance value always yields a smaller number of clusters. However, this decision involves a trade-off with respect to relaxing the constraint of cohesion in the cluster memberships \cite{chong2013efficient, chong2017automatic, chong2015analyzing}. As such, in this work, we attempt to determine the optimal total number of clusters by dividing the total number of classes with the following divisors : 5, 7, 10, 20, and 25. The numbers were chosen based on the package distribution of the ground truth that we generated and depends on the number of classes of the analysed software. \new{In this paper, we choose to use these divisors instead of an exhaustive approach to save computation time and obtain a range of the optimal number of classes. In practice, E-SC4R would allow the user to specify the number of clusters or divisors.}

We use different configurations of clustering algorithms which differ between the combination of different distance metrics, linkage algorithms, and the number of clusters. We then record the clustering results of each combination of the clustering algorithm on each version of the software, \new{to be compared with the ground truth}. For example: 

\new{
\begin{itemize}
    \item Agglomerative Hierarchical Configuration 1
    \begin{itemize}
        \item Linkage = Single
        \item Distance Metric = Euclidean
        \item Cluster Divisor = 5
    \end{itemize}
    \item Agglomerative Hierarchical Configuration 2
    \begin{itemize}
        \item Linkage = Complete
        \item Distance Metric = Cosine
        \item Cluster Divisor = 7
    \end{itemize}
\end{itemize}
}

\paragraph{Evaluation of clustering results}
As mentioned earlier, creating a reference model to act as the ground truth by engaging domain experts is expensive in terms of time and effort. On the other hand, the reliability of the package structure of the analysed software is strongly dependent on the experience of the software developers, as well as maturity of the analysed project. 

\new{In this paper, we create the reference model (ground truth) by looking at the most commonly occurring directory structure patterns for the 10 previous releases of the analysed software. A more detailed example is available in Section 3.4.}

The generated ground truth will then be used to compare against the clustering results that are produced by each of the hierarchical clustering algorithms. A more detailed example is available in Section \ref{groundtruth}.

To evaluate the performance of each hierarchical clustering algorithm against the reference model, we use MoJoFM metric proposed in the work by \cite{tzerpos1999mojo,wen2003optimal}. The MoJo family of metrics were widely used in the domain of software clustering to evaluate the performance of different clustering algorithms \cite{maqbool2007hierarchical,chong2017automatic, beck2016identifying, naseem2019euclidean}. Hence, in the remaining of this paper, the term performance of clustering algorithm refers to the MoJoFM value computed when comparing between the produced clustering results and the ground truth. 

\subsection{Bunch Clustering Algorithm}

Bunch supports three main clustering algorithms, namely hill-climbing algorithm, exhaustive clustering algorithm, and genetic algorithm. The authors claimed that exhaustive clustering algorithm only works well for small systems \cite{mitchell2002heuristic}, and the hill-climbing algorithm performs well for most software systems \cite{mitchell2006}. 

Bunch starts off by generating a random partition of the MDG. Then, depending on the chosen clustering algorithm (hill climbing, genetic algorithm, or exhaustive), it will cluster each of the random partitions in the population and select the result with the largest Modularisation Quality (MQ) as the suboptimal solution. MQ measures the quality of an MDG partition by taking into consideration the trade-off between the dependencies between the clustering entities (classes) of two distinct clusters (package/subsystem), and the dependencies between the clustering entities (classes) of the same cluster (package/subsystem). The assumption made is that high quality software systems should be designed with cohesive subsystems that are loosely coupled between each other. As the size of the problem (software system) increases, the probability of finding a good sub-optimal solution (MQ) also increases.

In this paper, we will be using a combination of different algorithms and MQ calculator to evaluate their performance against the chosen datasets. For example,
\new{
\begin{itemize}
    \item Bunch Configuration 1
    \begin{itemize}
        \item Algorithm = HillClimbing
        \item Calculator = TurboMQ
    \end{itemize}
    \item Bunch Configuration 2
    \begin{itemize}
        \item Algorithm = GeneticAlgorithm
        \item Calculator = TurboMQIncr
    \end{itemize}
\end{itemize}
}

\subsection{Dataset Collection}\label{datacollect}

For each chosen project, we compare the clustering results across 10 releases to ensure the stability of the clustering algorithm. Stability in software clustering is defined as the sensitivity of a particular clustering algorithm toward the changes in the dataset \cite{harman2005}. For any good clustering algorithm, small changes in the target software (clustering algorithm applied on multiple small increment releases of the same software) should not alter the clustering results significantly. Due to the way how we create the ground truth, 10 prior releases of the examined software will be needed to identify the common directory structure, as shown in Figure~\ref{fig:ground}. As such, 21 releases of the chosen project are required to form one of our selection criteria. 

Once we identified the suitable software, 21 versions of the project source code were then downloaded using GitHub CLI to a server based on the project link, release tag, and version name. In total, we had chosen 30 Java-based projects collected from GitHub as we could not find more which suited our search criteria. The selected projects are shown in Table \ref{table:listproject} and the complete set of datasets can be found on our Github page.\footnote{https://github.com/alvintanjianjia/SoftwareRemodularization}

\begin{table*}
    \caption{\new{List of chosen projects and their versions.}}
    \centering
    \begin{tabular}{|c|c|c|}
        \hline
    	Project Name & firstRelease & lastRelease \\
    	\hline
    	bkromhout-realm-java & v0.87.1 & v0.89.0 \\
    	\hline
    	btraceio-btrace & v1.3.4 & v1.3.9 \\
    	\hline
    	bytedeco-javacpp & 1.4 & 1.5.3 \\
    	\hline
    	codecentric-spring-boot-admin & 1.4.5 & 1.5.7 \\
    	\hline
    	codenvy-legacy-che-plugins & 3.13.4.4 & 3.9.5 \\
    	\hline
    	coobird-thumbnailator & 0.4.10 & 0.4.9 \\
    	\hline
    	dropwizard & v2.0.11 & v2.0.9 \\
    	\hline
    	dropwizard-metrics & v4.1.10.1 & v4.1.9 \\
    	\hline
    	evant-gradle-retrolambda & v3.3.0-beta1 & v3.7.0 \\
    	\hline
    	facebook-android-sdk & 5.15.1 & 5.9.0 \\
    	\hline
    	facebook-java-business-sdk	 & v3.2.7 & v3.3.6 \\
    	\hline
    	facebook-fresco & v1.14.2 & v1.9.0 \\
    	\hline
    	facebook-litho & v2020.05.11 & v2020.07.20 \\
    	\hline
    	facebook-react-native-fbsdk & 0.6.1 & v0.10.3 \\
    	\hline
    	google-cdep	 & 0.8.27 & 0.8.9 \\
    	\hline
    	google-dagger & 2.28.2 & 2.9.0 \\
    	\hline
    	google-error-prone & v2.1.1	 & v2.4.0 \\
    	\hline
    	google-gitiles & v0.2-9 & v0.4 \\
    	\hline
    	google-openrtb & 1.5.11 & 1.5.9 \\
    	\hline
    	google-openrtb-doubleclick & 1.5.14 & 1.5.9 \\
    	\hline
    	grpc-grpc-java & v1.30.1 & v1.9.1 \\
    	\hline
    	havarunner-havarunner & 0.8.4 & 0.9.5 \\
    	\hline
    	immutables-immutables & 2.6.0 & 2.7.5 \\
    	\hline
    	ionic-team-capacitor & 1.0.0 & 1.5.2 \\
    	\hline
    	jankotek-mapdb & 3.0.0 & 3.0.8 \\
    	\hline
    	javafunk-funk & 0.1.24 & 0.2.0 \\
    	\hline
    	javaparser & 3.6.9 & 3.13.10 \\
    	\hline
    	permissions-dispatcher & 2.1.0 & 2.4.0 \\
    	\hline
    	pxb1988-dex2jar & 0.0.9.14 & 0.0.9.9 \\
    	\hline
    	web3j & v4.5.2 & v4.6.1 \\
    	\hline
    \end{tabular}
    \label{table:listproject}
\end{table*}

%
%

The columns \textit{firstRelease} and \textit{lastRelease} in Figure~\ref{table:listproject} indicate the versions of the software that we examined and used for our experiments. Note that we use 21 incremental releases in between the stated \textit{firstRelease} and \textit{lastRelease} to ensure the stability of the clustering algorithms, and to generate the ground truth.

\subsection{Generation of Ground Truth}\label{groundtruth}

In this work, we attempt to improve the existing ground truth generation method by looking into the evolution of the analysed software over multiple releases \new{instead of the latest version package structure}. The creation of the ground truth is done via extraction of common directories across 10 previous releases of the software. For each version of the software, the previous 10 releases of the software were analysed and only the common file directory structure across all 10 versions will be extracted. 

\new{
Given the scale of the 30 open source projects over multiple releases, it is challenging to find domain experts for each of the software systems, which is a similar problem encountered in the work by \cite{naseem_deris_maqbool_li_shahzad_shah_2017}. Ground truths that are generated and manually approved by senior developers working on the project would only be practical for a handful of projects. However, this approach will not provide our research with enough data points for footprint visualisations during the evaluation of clustering results. Given that the current factual architecture of the system has been created by the open source developers or administrators themselves, it is reasonable to assume the current package structure is held to a certain standard \cite{beck2013impact} \cite{Alsarhan2020}. Additionally, these open source projects are among the highest rated Java projects on GitHub based on stars, which provides additional confidence on the correctness of the generated ground truth based on the evolution of the package structure over multiple releases.
}

 \begin{figure}[!ht]
 \centering
\includegraphics[width=75mm]{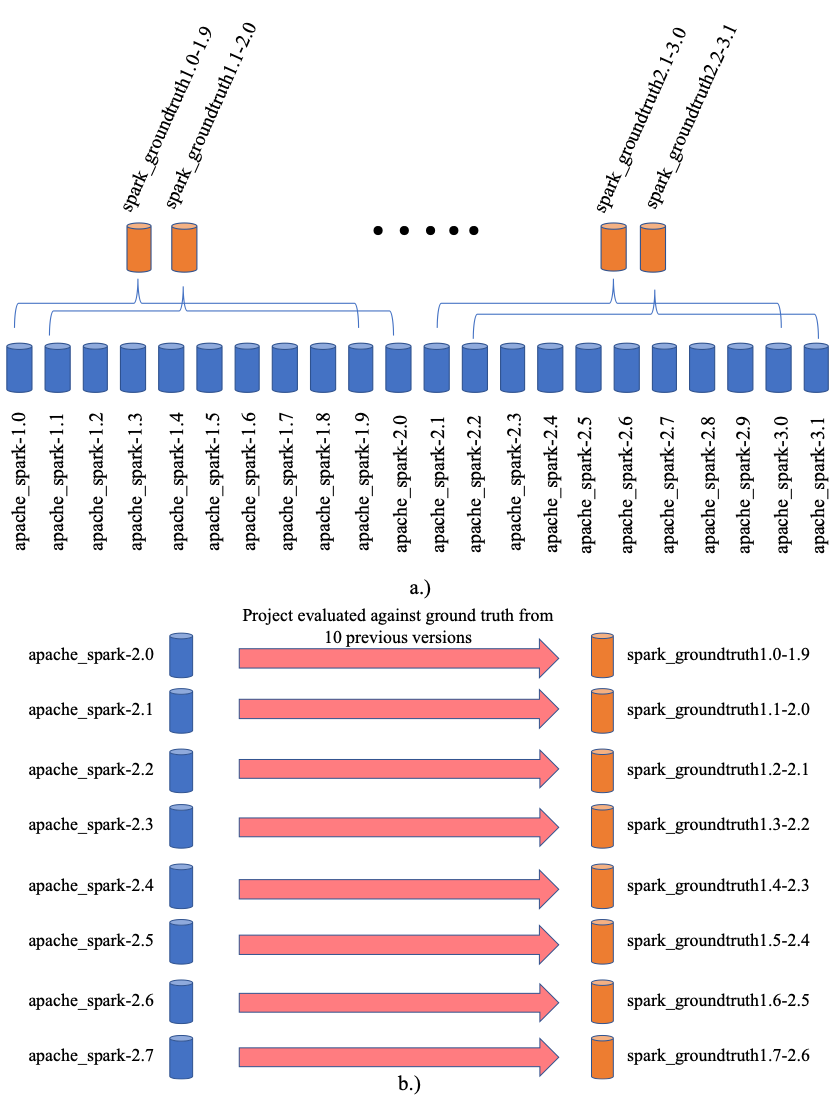}
\caption{\normalsize a.) Method used to select project releases and generate ground truth b.) Method used to evaluate clustering results against ground truth.\label{fig:ground}}
 \end{figure}
 
\newer{The package/directory structure for each selected software system is generated using an automated script, where it crawls the file structure of the GitHub repository. As such, we ensure that none of the selected test subjects are private repositories.} In the example shown in Figure~\ref{fig:ground}a, we use the common directories across apache\_spark-1.0 to apache\_spark-1.9 to generate the ground truth. Subsequently, this ground truth comprising the common directory from releases 1.0 to 1.9 will be used to evaluate against the clustering results that we produce in the next release, which is apache\_spark-2.0, as shown in Figure~\ref{fig:ground}b. To illustrate another simple example, given the following directory structure of a software in 3 incremental releases:

\begin{enumerate}
    \item V 2.5: src/var/c.java, src/var/d.java, tmp/eg/z.java
    \item V 2.6: src/var/c.java, src/var/d.java, temp/util/z.java
    \item V 2.7: src/var/c.java, src/var/d.java, temp/eg/z.java
\end{enumerate}

Based on our approach, only src/var/c.java and src/var/d.java are extracted from the given 3 versions. A parent-child cluster relationship would be defined based on the extracted directory paths, given by the parent \textit{contains} child.

For example, given that the following directory paths are extracted
\begin{enumerate}
    \item \textit{user/spark/java/a.java} 
    \item \textit{user/spark/java/b.java} 
    \item \textit{user/spark/main/test.java} 
\end{enumerate}

The following parent-child clusters will be created.
\begin{itemize}
    \item java \textit{contains} a.java
    \item java \textit{contains} b.java
    \item main \textit{contains} test.java
    \item spark \textit{contains} java
    \item spark \textit{contains} main
    \item user \textit{contains} spark
    
\end{itemize}



The final clustering result obtained is then taken as the ground truth, which will be used as the reference model when compared using the MoJoFM metric.

\subsection{Measuring Dependencies between Classes}

As mentioned earlier, for hierarchical clustering algorithms, we use \textit{Depends} \cite{jin2019enre} to quantify the strength of dependencies between classes in the examined software. The tool can create a $NxN$ matrix to show the types and frequency of dependencies between all the classes in the analysed software ($N$ = number of classes).

On the other hand, the Bunch tool uses Module Dependency Graph (MDG) to measure the strength of dependencies between classes\cite{mancoridis1998,mitchell2006}.

\subsection{Selection and Permutation of Chosen Clustering Algorithms}\label{exhaustive_testing_experiment_design}

For each of the chosen software projects, we ran different permutation of hierarchical and Bunch clustering algorithms based on the configuration shown in \new{Table \ref{table:parameter1} and Table \ref{table:parameter2}.}

\begin{table}[!h]
    \caption{\new{Summary of parameters and settings associated with hierarchical clustering.}}
    \centering
    \begin{tabular}{|p{3cm}|p{3cm}|}
        \hline
    	\hfil Parameters &  \hfil Values \\
    	\hline
    	\multirow{3}{6em}{Linkage Method} & \hfil Average \\
    	& \hfil Complete\\
    	& \hfil Single\\
    	\hline
    	\multirow{3}{6em}{Distance Metric} & \hfil Euclidean \\
    	& \hfil Cosine \\
    	& \hfil Manhattan \\
    	\hline
    	\multirow{3}{6em}{Divisor} & \hfil 5 \\
    	& \hfil 7 \\
    	& \hfil 10 \\
    	& \hfil 20 \\
    	& \hfil 25 \\
    	\hline
	\end{tabular}
    \label{table:parameter1}
\end{table}

\begin{table}[h!]
	\caption{\new{Summary of parameters and settings associated with Bunch clustering.}}
	\begin{center}
	\begin{tabular}{|p{3cm}|p{3cm}|}
	    \hline
	    \hfil Parameters & \hfil Values \\
		\hline
		\multirow{3}{6em}{Bunch Algorithm} & \hfil HillClimbing \\
    	& \hfil GeneticAlgorithm \\
    	& \hfil Exhaustive \\
		\hline
		\multirow{3}{6em}{Bunch Calculator} & \hfil TurboMQIncrW \\
    	& \hfil TurboMQIncr \\
    	& \hfil TurboMQ \\
    	& \hfil TurboMQW \\
	    & \hfil BasicMQ \\
	    \hline
	\end{tabular}
    \end{center}
\label{table:parameter2}
\end{table}

In total, we have 45 unique configurations of hierarchical clustering algorithms and 15 unique configurations of Bunch clustering algorithms to run on each chosen project. Furthermore, for each configuration, we ran it against the 10 prior releases of the target software to ensure the stability of the algorithm.

\subsection{Evaluation of Clustering Results}\label{mojo_evaluation}

\new{The clustering results are compared with the ground truth using MoJoFM} \cite{wen2003optimal}. Due to the size of the table, we are unable to show the full set of clustering results from all the chosen projects. The complete set of results can be assessed from our GitHub page. The summarised version of the clustering results, showing the top 10 clustering results for hierarchical clustering and Bunch are shown in \new{Table~\ref{table:top10agglo} and Table~\ref{table:top10bunch}} respectively.

\newer{
\subsection{Strengths and Weaknesses of Clustering Techniques}\label{ga_process_subsection}

\begin{table}[!hbt]
\centering
\caption{\newer{MoJo performance of Agglomerative clustering algorithm on Web3j version 4.5.2.}}
\begin{tabular}{ |c|c|c| } 
 \hline
 \textbf{Software} & \textbf{MoJoFM} & \textbf{Algorithm} \\ 
 \hline
 Web3j v4.5.2 & 95.56 & Cosine Average 10 \\
 \hline
 Web3j v4.5.2 & 94.86 & Cosine Average 7 \\
 \hline
 Web3j v4.5.2 & 91.83 & Cosine Average 5 \\
 \hline
\end{tabular}
\label{table:mojo_performance_metric}
\end{table}

\begin{table*}[!hbt]
    \centering
    \caption{\newer{Example of software feature set used during GA.}}
    \begin{tabular}{|c|c|c|c|c|c|}
    \hline
    Software & LOC & Total Number of Methods & Number of StaticMethods & DIT & Most Suitable Algorithm\\
    \hline
    Web3j v4.5.2 & 1565 & 123 & 33 & 38 & Cosine Average 10\\
    \hline
    Web3j v4.5.3 & 1678 & 125 & 32 & 38 & Cosine Average 7\\
    \hline
    \end{tabular}
    \label{tab:ga_dataset}
\end{table*}

In this subsection, we illustrate with examples, how the strength and weakness of clustering techniques are identified. Table \ref{table:mojo_performance_metric} shows an example of the results that we obtained during the experiment phase from Sections \ref{exhaustive_testing_experiment_design} and \ref{mojo_evaluation}. Since a higher MoJoFM value indicates a higher suitability of the clustering algorithm, we will assign "Cosine Average 10" as the most suitable algorithm for Web3j version 4.5.2. This is done for all the selected software systems and their releases.

Table \ref{tab:ga_dataset} shows some example sets of software features (LOC, Total Number of Methods, Number of StaticMethods, DIT) of the selected test subjects, and the most suitable algorithm identified during the experiment phase. During the GA process, we would select a subset of software features and evaluate the classification accuracy. For example:
\begin{itemize}
    \item Subset 1: (LOC, Total Number of Methods, Number of StaticMethods) in predicting most suitable algorithm
    \item Subset 2: (Total Number of Methods, Number of StaticMethods, DIT) in predicting most suitable algorithm
\end{itemize}
}

\begin{table*}[h!]
	\caption{Top 10 hierarchical clustering results based on MoJoFM.}
	\begin{tabular}{|c|c|c|c|c|c|}
		\hline
		Project Name & Project Version & Divisor & Affinity & Linkage & MoJoFM \\
		\hline
		facebook-facebook-java-business-sdk & 3.2.7 & 25 & euclidean & single & 98.657 \\
		\hline
		facebook-facebook-java-business-sdk & 3.2.8 & 25 & euclidean & single & 96.979 \\
		\hline
		facebook-facebook-java-business-sdk & 3.2.9 & 25 & euclidean & single & 96.812 \\
		\hline
		facebook-facebook-java-business-sdk & 3.3.1 & 25 & euclidean & single & 96.644 \\
		\hline
		facebook-facebook-java-business-sdk & 3.3.5 & 25 & euclidean & single & 96.643 \\
		\hline
		facebook-facebook-java-business-sdk & 3.3.6 & 25 & euclidean & single & 96.642 \\
		\hline
		facebook-facebook-java-business-sdk & 3.3.0 & 25 & euclidean & single & 96.476 \\
		\hline
		facebook-facebook-java-business-sdk & 3.3.2 & 25 & euclidean & single & 96.475 \\
		\hline
		facebook-facebook-java-business-sdk & 3.3.3 & 25 & euclidean & single & 96.474 \\
		\hline
		javafunk-funk & 0.2.0 & 25 & euclidean & single & 96.268 \\
		\hline
	\end{tabular}
\label{table:top10agglo}
\end{table*}

\begin{table*}[h!]
	\caption{Top 10 Bunch clustering results based on MoJoFM.}
	\begin{tabular}{|c|c|c|c|c|}
		\hline
		Project Name & Project Version & Algorithm & Calculator & MoJoFM \\
		\hline
		google-openrtb & 1.5.7 & GA & TurboMQIncrW & 90.909 \\
		\hline
		google-openrtb & 1.5.5 & HillClimbing & BasicMQ & 90.909 \\
		\hline
		facebook-react-native-fbsdk & 0.10.2 & Exhaustive & TurboMQIncrW & 90.476 \\
		\hline
		facebook-react-native-fbsdk & 0.10.0 & HillClimbing & TurboMQ & 90.476 \\
		\hline
		bytedeco-javacpp & 1.4.2 & HillClimbing & TurboMQW & 88.961 \\
		\hline
		google-openrtb & 1.5.9 & Exhaustive & TurboMQIncr & 88.637 \\
		\hline
		google-openrtb & 1.5.2 & Exhaustive & TurboMQIncr & 88.637 \\
		\hline
		google-openrtb & 1.5.4 & HillClimbing & BasicMQ & 88.636 \\
		\hline
		google-openrtb & 1.5.6 & Exhaustive & TurboMQ & 88.636 \\
		\hline
		google-openrtb & 1.5.11 & HillClimbing & BasicMQ & 86.363 \\
		\hline
	\end{tabular}
\label{table:top10bunch}
\end{table*}

\section{Results and Discussion}
The proposed E-SC4R framework \new{identifies the} most significant software features, which have an impact on the performance of clustering techniques. The resulting SVM predictions are plotted in the reduced instance space as shown in Figure~\ref{fig:combined_algorithm_spread}.

Based on the output of the SVM model, we took a deeper dive into the accuracy, precision, and recall scores of each algorithm, and found out that most of the algorithms with distinct separable clusters are with high recall scores, while the algorithms with indistinguishable clusters are with low recall scores.

\new{
Distinct separable clusters are manually identified from the footprint visualisations, where for a specific range of software metrics, the algorithm performs undoubtedly the best for clustering these software systems. Drawing an example from Figure~\ref{fig:combined_algorithm_spread}, in the range where 0.2 < z1 < 0.4, and 0.6 < z2 < 0.8, cosine\_average\_10 performs the best for these software systems. The values that z1 and z2 represent are explained in more detail under Section \ref{sec:combined_pca_footprint_visualization}. }

Given that,
\begin{itemize}
    \newer{
    \item True Positive = Given a new software system, E-SC4R correctly identifies a suitable algorithm
    \item True Negative = Given a new software system, E-SC4R correctly rejects an unsuitable algorithm
    \item False Positive = Given a new software system, E-SC4R identifies an algorithm which is unsuitable
    \item False Negative = Given a new software system, E-SC4R rejects a suitable algorithm}
    \item Accuracy = The number of times E-SC4R is able to accurately predict the \newer{most suitable} clustering algorithm or reject the wrong clustering algorithm out of the total predictions made: \( \frac{TP + TN}{TP + TN + FP + FN} \)
    \item Precision = The number of times E-SC4R is accurate in predicting the \newer{most suitable} algorithm out of all the times the algorithm is predicted by E-SC4R: \( \frac{TP}{TP + FP} \)
    \item Recall = The number of times E-SC4R is accurate in predicting the \newer{most suitable} algorithm out of all the times the best algorithm should have been predicted : \( \frac{TP}{TP + FN} \)
    
\end{itemize}

We are able to identify 3 main patterns/clusters of algorithms from the results shown in Table \ref{table:svm_cluster_type}. Note that due to the size of the table, we only show some of the examples in the last column of Table \ref{table:svm_cluster_type}. The information about the average accuracy, precision, and recall of agglomerative and Bunch algorithm produced from the SVM model (on the 300 releases) are shown in Tables  \ref{table:svm_agglo_cosine_accuracy_precision_recall},
\ref{table:svm_agglo_euclidean_accuracy_precision_recall},
\ref{table:svm_agglo_manhattan_accuracy_precision_recall}, and
\ref{table:svm_bunch_accuracy_precision_recall}.

\begin{table*}[h!]
	
	\caption{\new{Categorisation of clustering results based on SVM.}}
	\begin{tabular}{p{0.05\linewidth}p{0.2\linewidth}p{0.35\linewidth}p{0.2\linewidth}}
	
		Cluster & Type & Observation & Examples\\
		\hline
		$c_1$ & High accuracy, high precision and high recall & We are able to correctly predict when these algorithms are suitable, as well as which features are the most important when determining the selection of these algorithms. & cosine\_single\_20, euclidean\_single\_25, manhattan\_single\_15, hillclimbing\_turbomqincrw \\
		\hline
        $c_2$ & Low accuracy, low precision and low recall & It is hard to predict whether these algorithms are suitable. & cosine\_complete\_20, euclidean\_complete\_15, manhattan\_complete\_15, euclidean\_average\_15, ga\_turbomq, ga\_turbomqincr \\
        \hline
        $c_3$ & High accuracy, high precision and low recall & The model is generally able to accurately predict when these algorithms are suitable, but most of the time the model prioritises other algorithms compared to the selected algorithms. & cosine\_average\_10, cosine\_single\_5, exhaustive\_turbomqincr, exhaustive\_turbomqincrw \\
        \hline
	\end{tabular}
\label{table:svm_cluster_type}
\end{table*}

Algorithms that fall into $c_1$ which possess high recall are preferred as we would be able to easily identify software features that contribute to determining whether a particular algorithm is the \newer{most suitable} for the given software systems.

\begin{table*}[h!]
	\caption{Performance of SVM model for agglomerative cosine algorithm.}
	\begin{tabular}{|l|r|r|r|}
	    \hline
		Algorithm & Accuracy & Precision & Recall\\
		\hline
		cosine\_average\_10 & 88.7 & 90.0 & 21.4 \\
		cosine\_average\_15 & 81.7 & 94.7 & 25.0 \\
		cosine\_average\_20 & 69.0 & 66.7 & 41.0 \\
		cosine\_average\_25 & 63.3 & 82.8 & 34.9 \\
		cosine\_average\_5 & 97.3 & 70.0 & 58.3 \\
		cosine\_average\_7 & 96.3 & 100.0 & 47.6 \\
		cosine\_complete\_10 & 93.7 & 90.0 & 33.3 \\
		cosine\_complete\_15 & 56.3 & 8.5 & 29.4 \\
		cosine\_complete\_20 & 64.3 & 20.0 & 34.0 \\
		cosine\_complete\_25 & 49.3 & 27.1 & 67.6 \\
		cosine\_complete\_5 & 97.0 & 70.0 & 53.8 \\
		cosine\_complete\_7 & 97.0 & 100.0 & 52.6 \\
		cosine\_single\_10 & 55.7 & 90.0 & 6.4 \\
		cosine\_single\_15 & 85.3 & 86.9 & 95.8 \\
		cosine\_single\_20 & 90.7 & 90.4 & 99.6 \\
		cosine\_single\_25 & 89.0 & 88.6 & 99.6 \\
		cosine\_single\_5 & 88.0 & 100.0 & 21.7 \\
		cosine\_single\_7 & 79.7 & 100.0 & 14.1 \\
        \hline
	\end{tabular}
\label{table:svm_agglo_cosine_accuracy_precision_recall}
\end{table*}

\begin{table*}[h!]
	\caption{Performance of SVM model for agglomerative Euclidean algorithm.}
	\begin{tabular}{|l|r|r|r|}
	    \hline
		Algorithm & Accuracy & Precision & Recall\\
		\hline
		euclidean\_average\_10 & 89.0 & 90.0 & 22.0 \\
		euclidean\_average\_15 & 51.7 & 27.0 & 54.7 \\
		euclidean\_average\_20 & 64.0 & 45.3 & 43.4 \\
		euclidean\_average\_25 & 58.7 & 54.4 & 99.3 \\
		euclidean\_average\_5 & 96.3 & 70.0 & 46.7 \\
		euclidean\_average\_7 & 94.0 & 100.0 & 35.7 \\
		euclidean\_complete\_10 & 88.3 & 90.0 & 20.9 \\
		euclidean\_complete\_15 & 63.3 & 23.3 & 17.9 \\
		euclidean\_complete\_20 & 59.7 & 49.2 & 52.9 \\
		euclidean\_complete\_25 & 59.0 & 55.0 & 98.7 \\
		euclidean\_complete\_5 & 95.3 & 70.0 & 38.9 \\
		euclidean\_complete\_7 & 95.0 & 100.0 & 40.0 \\
		euclidean\_single\_10 & 75.0 & 90.0 & 10.8 \\
		euclidean\_single\_15 & 67.0 & 65.9 & 100.0 \\
		euclidean\_single\_20 & 83.0 & 82.4 & 100.0 \\
		euclidean\_single\_25 & 87.7 & 87.2 & 100.0 \\
		euclidean\_single\_5 & 93.3 & 70.0 & 29.2 \\
		euclidean\_single\_7 & 85.7 & 100.0 & 18.9 \\
		
        \hline
	\end{tabular}
\label{table:svm_agglo_euclidean_accuracy_precision_recall}
\end{table*}

\begin{table*}[h!]
	\caption{Performance of SVM model for agglomerative Manhattan algorithm.}
	\begin{tabular}{|l|r|r|r|}
	    \hline
		Algorithm & Accuracy & Precision & Recall\\
		\hline
		manhattan\_average\_10 & 70.3 & 90.0 & 9.3 \\
		manhattan\_average\_15 & 84.0 & 83.0 & 99.1 \\
		manhattan\_average\_20 & 92.7 & 92.1 & 100.0 \\
		manhattan\_average\_25 & 94.0 & 93.6 & 100.0 \\
		manhattan\_average\_5 & 92.3 & 70.0 & 25.9 \\
		manhattan\_average\_7 & 87.0 & 100.0 & 20.4 \\
		manhattan\_complete\_10 & 85.7 & 90.0 & 17.6 \\
		manhattan\_complete\_15 & 59.3 & 35.0 & 48.8 \\
		manhattan\_complete\_20 & 59.0 & 54.8 & 99.3 \\
		manhattan\_complete\_25 & 77.7 & 75.6 & 99.5 \\
		manhattan\_complete\_5 & 95.0 & 60.0 & 35.3 \\
		manhattan\_complete\_7 & 95.0 & 100.0 & 40.0 \\
		manhattan\_single\_10 & 62.3 & 90.0 & 7.4 \\
		manhattan\_single\_15 & 93.3 & 92.9 & 100.0 \\
		manhattan\_single\_20 & 92.7 & 92.1 & 100.0 \\
		manhattan\_single\_25 & 94.0 & 93.6 & 100.0 \\
		manhattan\_single\_5 & 90.0 & 70.0 & 20.6 \\
		manhattan\_single\_7 & 76.7 & 100.0 & 12.5 \\
        \hline
	\end{tabular}
\label{table:svm_agglo_manhattan_accuracy_precision_recall}
\end{table*}

\begin{table*}[hbt!]
	\caption{Performance of SVM model for Bunch algorithm.}
	\begin{tabular}{|l|r|r|r|}
	    \hline
		Algorithm & Mean Accuracy & Mean Precision & Mean Recall\\
		\hline
		exhaustive\_basicmq & 60.3 & 11.5 & 24.4 \\
		exhaustive\_turbomq & 82.0 & 0.0 & 0.0 \\
		exhaustive\_turbomqincr & 79.0 & 10.0 & 7.7 \\
		exhaustive\_turbomqincrw & 77.0 & 19.0 & 7.1 \\
		exhaustive\_turbomqw & 82.0 & 0.0 & 0.0 \\
		ga\_basicmq & 70.3 & 19.6 & 20.0 \\
		ga\_turbomq & 85.7 & 0.0 & 0.0 \\
		ga\_turbomqincr & 82.7 & 0.0 & 0.0 \\
		ga\_turbomqincrw & 72.3 & 22.8 & 25.0 \\
		ga\_turbomqw & 61.0 & 13.3 & 23.5 \\
		hillclimbing\_basicmq & 73.3 & 22.7 & 17.9 \\
		hillclimbing\_turbomq & 73.7 & 26.5 & 23.2 \\
		hillclimbing\_turbomqincr & 79.7 & 16.7 & 12.2 \\
		hillclimbing\_turbomqincrw & 51.0 & 21.6 &63.6 \\
		hillclimbing\_turbomqw & 85.7 & 0.0 & 0.0 \\
        \hline
	\end{tabular}
\label{table:svm_bunch_accuracy_precision_recall}
\end{table*}


\subsection{PCA Visualisation}
To visualise the results in a meaningful way, we apply PCA as a dimensionality reduction technique on the optimal subset of software features. The aim is to plot the performance of the different clustering algorithms across the project space in 2D, which is likely to reveal where the clustering algorithms perform well, and where are their weaknesses. Two new axes were created, which are linear combinations of the selected set of software features. \textbf{Projecting it using the two principal components holds 85\% of the variation in the data.}  

\subsubsection{Combined PCA Visualisation of Agglomerative and Bunch algorithms} \label{sec:combined_pca_footprint_visualization}
\begin{figure*}[!ht]
 \centering
\includegraphics[width=0.55\linewidth]{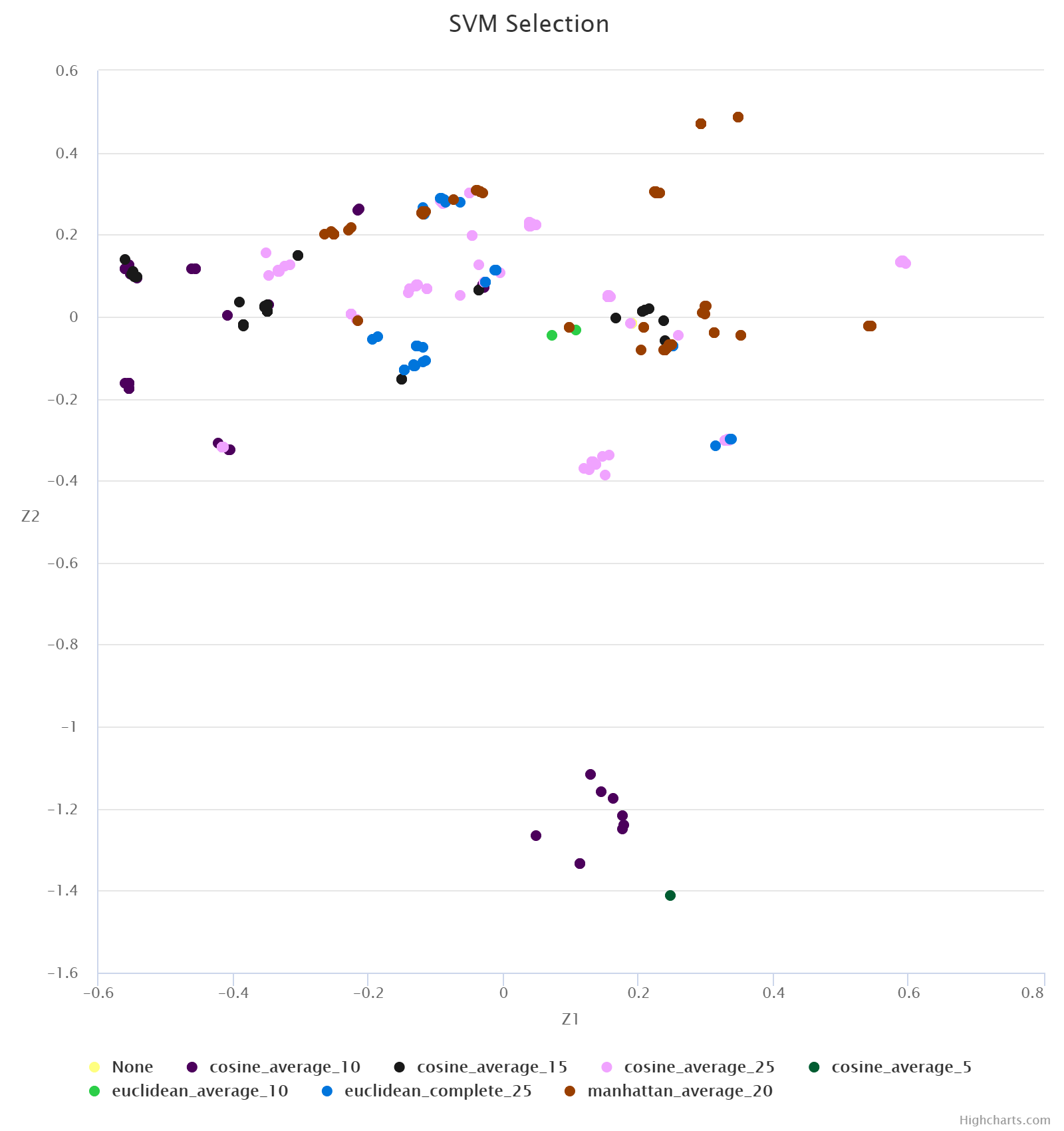}
\caption{\normalsize  Combined algorithm spread.\label{fig:combined_algorithm_spread}}
\end{figure*}

A combined visualisation Figure \ref{fig:combined_algorithm_spread} is created to have a general overview on the spread of the algorithm and features. \newer{A dot in Figure \ref{fig:combined_algorithm_spread} represents a software system. The color represents which is the most optimum clustering algorithm and its configuration.} Based on the MoJoFM results, an initial comparison was made on which algorithms are to be prioritised among the 300 projects (30 unique projects with 10 releases each), where some examples of the projects are shown in Table~ \ref{table:listproject}. Recall that for each of the chosen 30 projects, we perform different configurations of agglomerative and Bunch clustering algorithms over 10 releases. The best performing algorithm (in terms of MoJoFM values) will be prioritised.

 The coordinate system that defines the new instance space is defined as

\begin{equation}
\small
\label{eq:combine}
\begin{bmatrix}
	z_1 \\
	z_2
\end{bmatrix}
= 
 \begin{bmatrix}
	{-0.1027} & {0.0546}  \\
	 {-0.0137} & {0.1251}  \\
	 {0.1056} & {-0.1107}   \\
	 {-0.0784} & {0.0747}   \\
	 {0.0708} & {0.0566}   \\
	 {0.091} & {-0.0484}   \\
	 {0.1015} & {0.0525}   \\
	 {-0.0418} & {-0.0474}  \\
	 {0.1215} & {0.0154}  \\
	 {0.0578} & {0.1086}  \\
\end{bmatrix}
 \begin{bmatrix}
	{\text{staticMethods sum}}  \\
	{\text{modifiers mean}} \\
	{\text{defaultMethods mean}} \\
	{\text{maxNestedBlocks mean}} \\
	{\text{totalMethods mean}} \\
	{\text{protectedMethods mean}} \\
	{\text{finalFields mean}} \\
	{\text{stringLiteralsQty mean}} \\
	{\text{lambdasQty mean}} \\
	{\text{returnQty mean}} \\
	
\end{bmatrix}
\end{equation}

As seen from the visualisation in Figure \ref{fig:combined_algorithm_spread}, agglomerative algorithms are heavily prioritised over Bunch algorithms, where most of the points of the instance space prioritises the usage of agglomerative algorithms. An interpretation of Figure 6 could be seen as such, where by looking at Equation \ref{eq:combine} and the visualisation generated, when $z_2$ is within the range of -1 to -1.4, and $z_1$ is within the range of 0 to 0.2, the algorithm that is prioritised is cosine\_average\_10. Based on Table \ref{table:compare_bunch_agglo} and Equation \ref{eq:combine}, we can see that agglomerative algorithms are prioritised over bunch algorithms if MoJoFM is used as the main evaluation metric based on the output from the SVM framework as well as the raw data from the first part of the experiments.

\new{A higher value for the individual feature in Equation~\ref{eq:combine} would mean that the feature has a higher influence on predicting which algorithm is the best, and lower values would mean a lower feature importance. For example, maxNestedBlocks mean in Equation~\ref{eq:combine} has comparatively low values. This means that the same clustering algorithm (with its corresponding configuration) may be suitable for programs with vastly different values for this feature.}

Upon further investigation, we discovered that although agglomerative clustering algorithm appears to be the superior algorithm when measured against MoJoFM, it usually generates many small clusters with few classes (5-10 classes). On the other hand, Bunch tends to generate clustering results with lesser number of clusters and more equal classes inside each cluster, which might make it easier for software maintainers to follow the suggested decomposition. Our findings largely agree with the experiments done by Wu et al. \cite{wu2005comparison} where they discovered that algorithms that give good
clustering results according to one criterion (i.e. MoJoFM) often do not give good results according to other criterion (i.e. size and number of clusters).

\begin{table}[h!]
\centering
\caption{Performance of agglomerative vs Bunch clustering algorithm over 300 projects.}
\begin{tabular}{ |c|c| } 
 \hline
 \textbf{Algorithms} & \textbf{Number of times algorithm is prioritised} \\ 
 \hline
 Agglomerative & 291 \\
 \hline
 Bunch & 9 \\ 
 \hline
\end{tabular}
\label{table:compare_bunch_agglo}
\end{table}

As such, we have decided to analyse the strengths and weaknesses of agglomerative and Bunch clustering algorithms separately using PCA, instead of combining the two.

\new{
\begin{tcolorbox}
\textbf{Summary of Combined PCA Visualisations:} 
Agglomerative algorithms are prioritised over bunch algorithms if MoJoFM is used as the main evaluation metric. Agglomerative algorithms usually generate many small clusters with few classes (5-10). Bunch algorithms usually generate fewer clusters with a more balanced spread of the number of classes inside each cluster.
\end{tcolorbox}
}

\subsection{Agglomerative PCA Visualisation}
\begin{figure*}[!ht]
    \centering
    \includegraphics[width=0.55\linewidth]{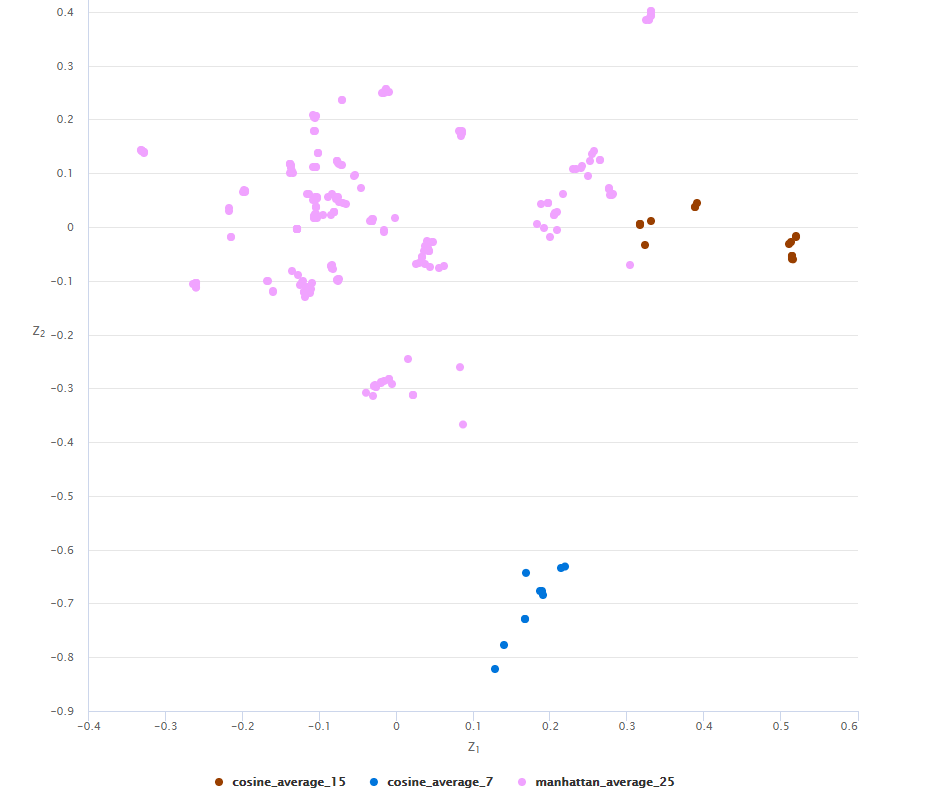}
    \caption{Agglomerative algorithm selection.}
    \label{fig:Agglo_svm_selection}
\end{figure*}

Figure \ref{fig:Agglo_svm_selection} illustrates the footprint visualisation generated for agglomerative algorithms. 

The coordinate system that defines the new instance space is defined as:
\newline

\begin{equation}
\small
\label{eq:agglo_pca_equation}
\begin{bmatrix}
	z_1 \\
	z_2
\end{bmatrix}
= 
 \begin{bmatrix}
	{0.0224} & {0.0178} \\
	 {-0.1018} & {0.0626} \\
	 {-0.0086} & {0.0677} \\
	 {-0.0459} & {-0.005} \\
	 {-0.0708} & {0.0803} \\
	 {0.0695} & {0.1282} \\
	 {0.0908} & {0.003} \\
\end{bmatrix}
 \begin{bmatrix}
	{\text{staticMethods std}}  \\
	{\text{privateMethods mean}}  \\
	{\text{subClassesQty mean}}  \\
	{\text{cbo mean}}  \\
	{\text{modifiers max}}  \\
	{\text{publicMethods mean}} \\
	{\text{anonymousClassQty mean}} \\
\end{bmatrix}
\end{equation}

where the footprint shows 3 main clusters.
\begin{itemize}
    \item cosine\_average\_15 (brown)
    \item cosine\_average\_7 (blue)
    \item manhattan\_average\_25 (pink)
\end{itemize}

The seven software features in Equation~\ref{eq:agglo_pca_equation} are identified to be the most impactful on the performance of the different clustering algorithms used for software remodularisation. \newer{A higher value for the individual feature in Equation~\ref{eq:agglo_pca_equation} would mean that the feature has a higher influence on predicting which algorithm is the best, and lower values would mean a lower feature importance. For example, std of staticMethods in Equation~\ref{eq:agglo_pca_equation} has comparatively low values. This means that the same clustering algorithm (with its corresponding configuration) may be suitable for programs with vastly different values for this feature.}

We noticed that six out of the seven software features that form the coordinate system in Figure \ref{fig:Agglo_svm_selection} are size metrics. This shows that for agglomerative clustering algorithms, size metrics have a stronger influence over the performance of the algorithm.

\subsubsection{Agglomerative PCA Relationship between Features and Clusters}
\begin{figure*}[!ht]
 \centering
 \subfigure[Footprints of Cosine Average 7.]{\includegraphics[width=0.47\linewidth]{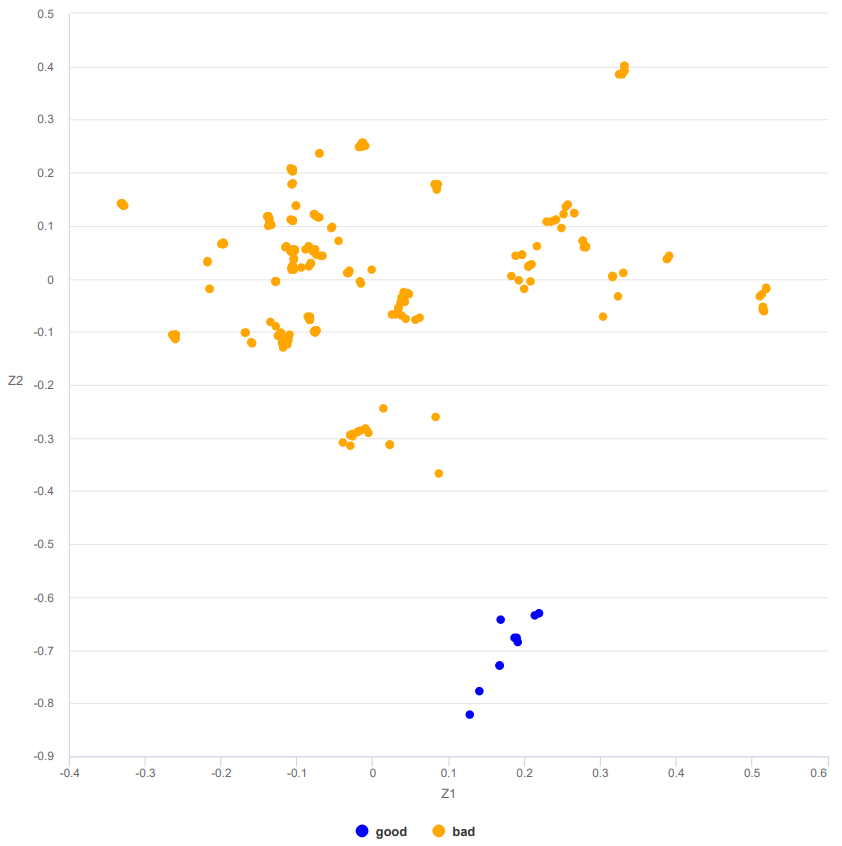}\label{fig:agglo_algo_cosine_average_7}}
 \subfigure[Footprints of Cosine Average 15.]{\includegraphics[width=0.47\linewidth]{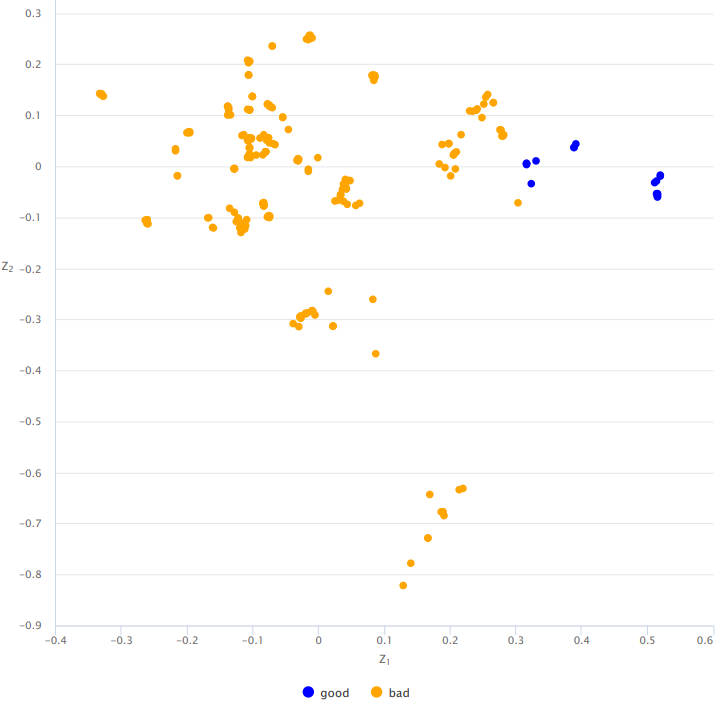}\label{fig:agglo_algo_cosine_average_15}}
 \subfigure[Footprints of Manhattan Average 25.]{\includegraphics[width=0.47\linewidth]{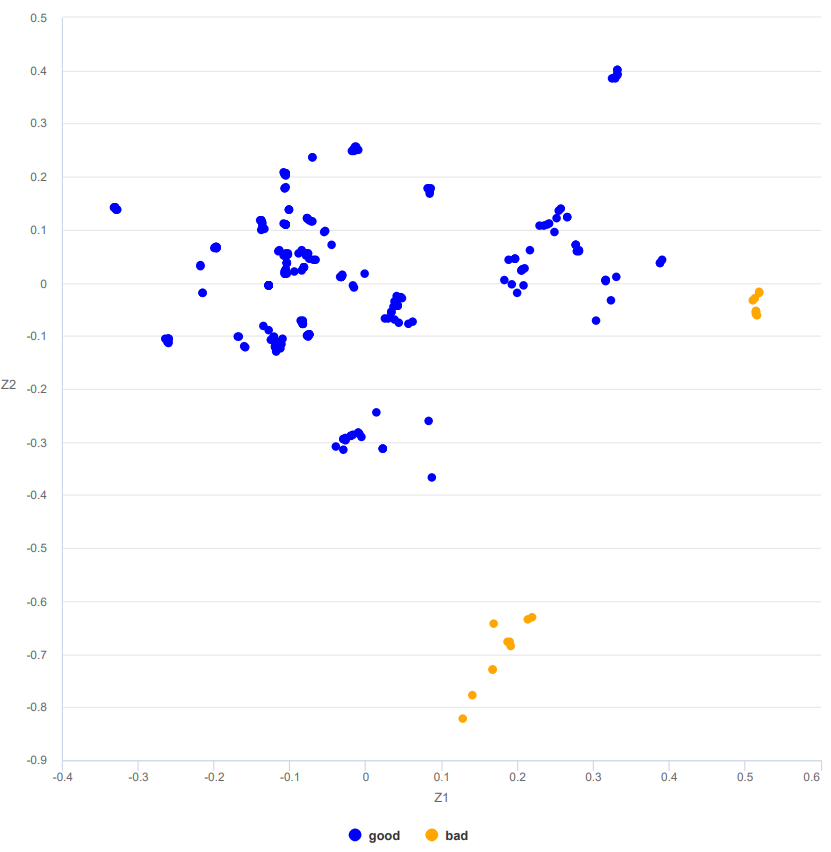}\label{fig:agglo_algo_manhattan_average_25}}\\
 \caption{Agglomerative algorithm footprint visualisation.\label{fig:footprint_agglo_algo}}
 \end{figure*}

\begin{figure*}[p!]
\centering
 \subfigure[Distribution of modifiers\_max.]{\includegraphics[width=0.47\linewidth]{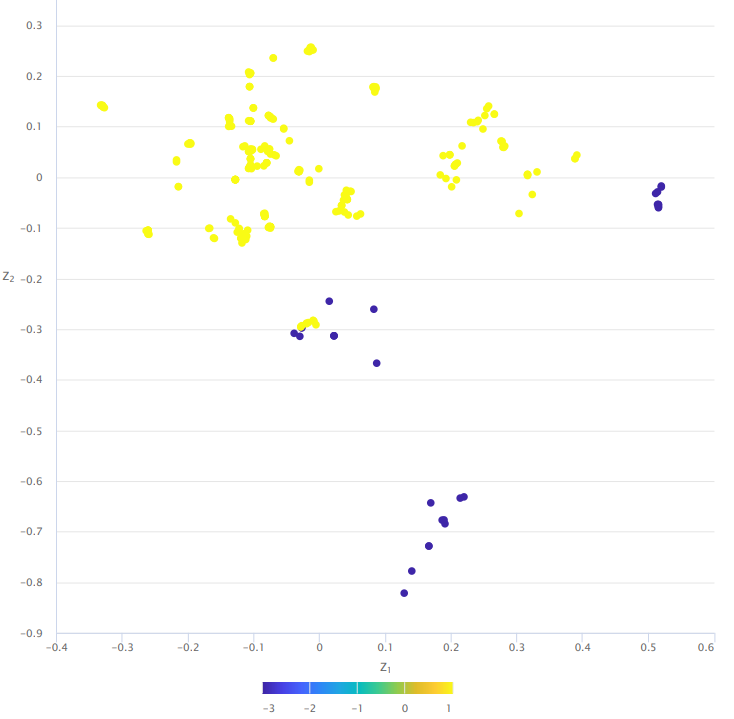}\label{fig:agglo_feature_modifiers_max}}
 \subfigure[Distribution of public methods\_mean.]{\includegraphics[width=0.47\linewidth]{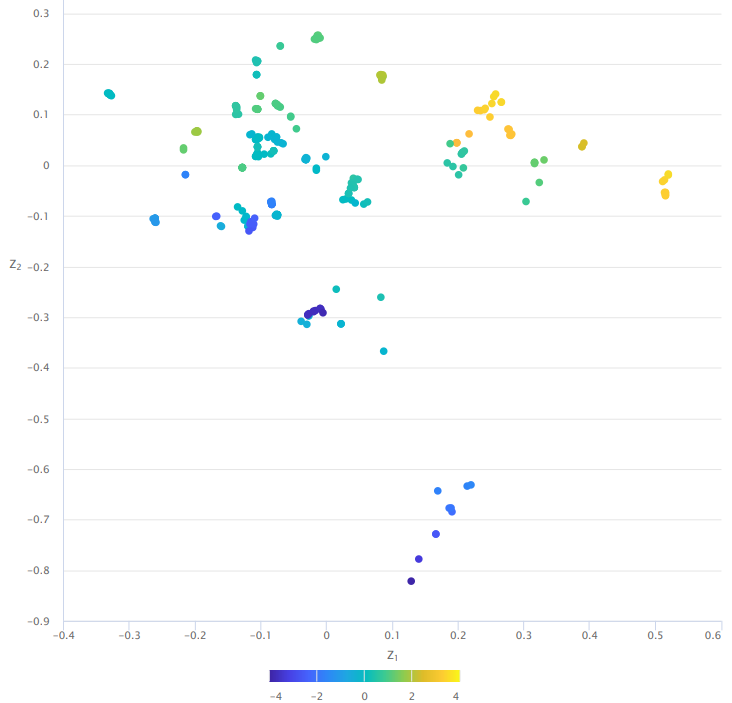}\label{fig:agglo_feature_publicMethods_mean}}
 \subfigure[Distribution of static methods\_std.]{\includegraphics[width=0.47\linewidth]{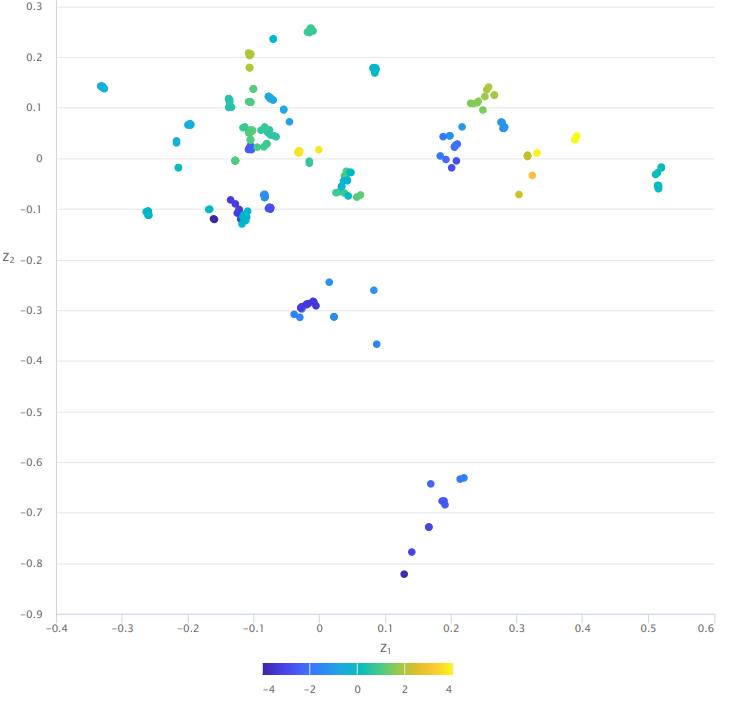}\label{fig:agglo_feature_staticMethods_std}}
 \subfigure[Distribution of private methods\_mean.]{\includegraphics[width=0.47\linewidth]{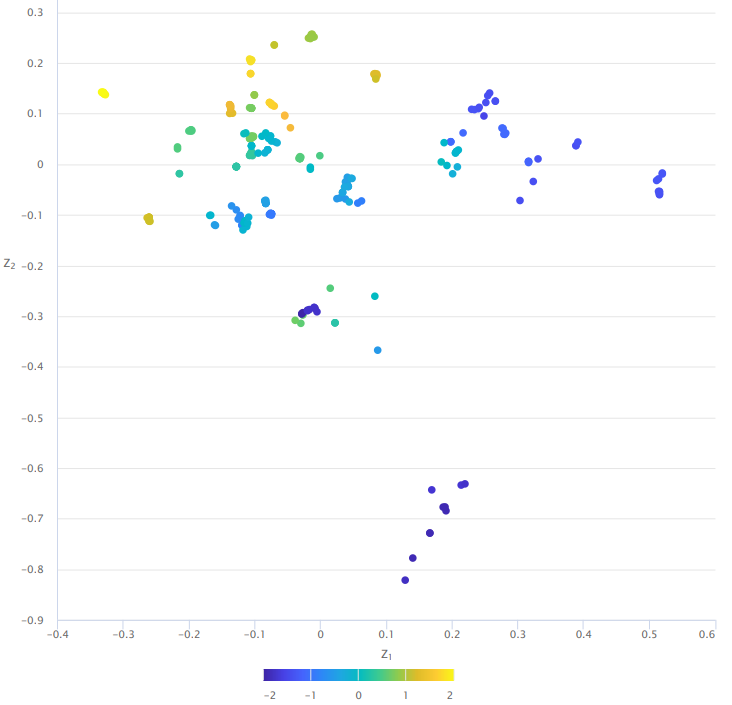}\label{fig:agglo_feature_privateMethods_mean}}
 \caption{Agglomerative software features footprint visualisation.\label{fig:footprint_agglo_features_a}}
 \end{figure*}

\begin{figure*}[!ht]
\centering
 \subfigure[Distribution of Coupling Between Objects\_mean.]{\includegraphics[width=0.47\linewidth]{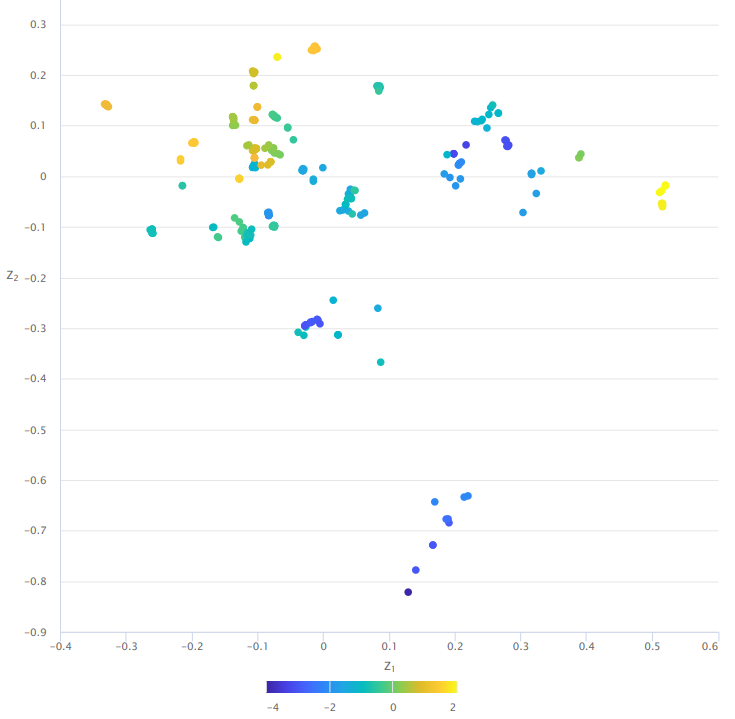}\label{fig:agglo_feature_cbo_mean}}
 \subfigure[Distribution of SubClassesQty\_mean.]{\includegraphics[width=0.47\linewidth]{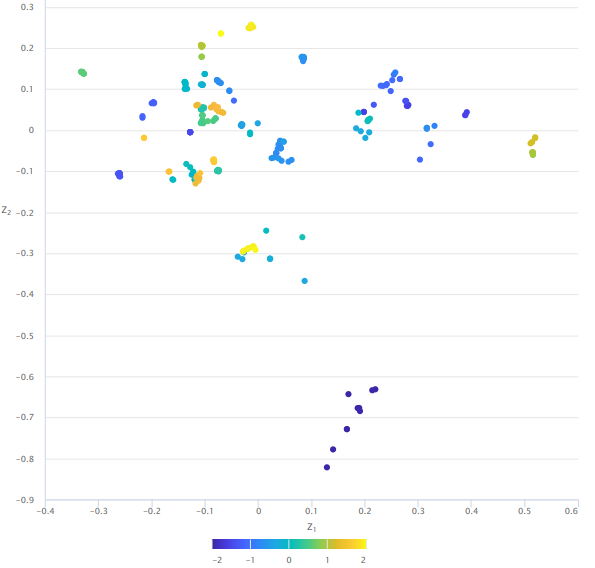}\label{fig:agglo_feature_subClassesQty_mean}}
 \subfigure[Distribution of AnonymousClassesQty\_mean.]{\includegraphics[width=0.47\linewidth]{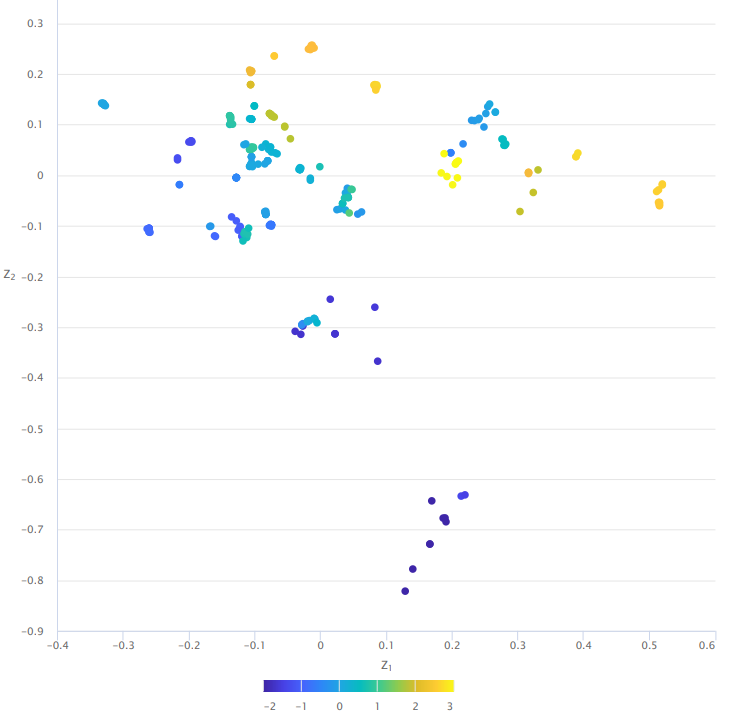}\label{fig:agglo_feature_anonymousClassesQty_mean}}
\caption{Agglomerative software features footprint visualisation.\label{fig:footprint_agglo_features_b}}
\end{figure*}

Using the new coordinate system for agglomerative clustering algorithm, we visualise the footprints of the different techniques as shown in Figures \ref{fig:footprint_agglo_algo}, \ref{fig:footprint_agglo_features_a} and \ref{fig:footprint_agglo_features_b}. We show the results for the prioritised clustering methods individually (Figures~\ref{fig:agglo_algo_cosine_average_7}-\ref{fig:agglo_algo_manhattan_average_25}), by setting the threshold of good performance if the quality of the software clustering is above 70\% (MoJoFM). Each data point represents a project, which is labelled as good if the performance of the MoJoFM score is above 70\%, and as bad otherwise.

To better understand why certain clustering algorithms work better for software projects in the cluster, we did a side-by-side comparison of software features and SVM model performance (Figures \ref{fig:footprint_agglo_features_a}, \ref{fig:footprint_agglo_features_b}) to try and draw correlations.

By comparing Figures \ref{fig:footprint_agglo_algo}, \ref{fig:footprint_agglo_features_a} and \ref{fig:footprint_agglo_features_b}, we are able to draw similarities between the algorithm's footprint patterns and the distribution of the features' values. This means that the following features are the most important when determining the priority of the algorithms.
\begin{itemize}
    \item Modifiers\_max - \textit{Figure \ref{fig:agglo_feature_modifiers_max}} and Manhattan Average 25 (linkage method; distance metric; cluster divisor) - \textit{Figure \ref{fig:agglo_algo_manhattan_average_25}}
    \begin{itemize}
        \item There are distinct distributions between the top left and bottom right clusters which are reflected in both footprints. Representing the software features of the target software \textbf{$s\in S$} in the new instance space, when the software features fall in the region of z\_2 > -0.2 and z\_1 < 0.4, modifiers\_max is the most important feature in determining whether Manhattan Average 25 is the \newer{most suitable} clustering algorithm.
    \end{itemize}
    \item Public Methods\_mean - \textit{Figure \ref{fig:agglo_feature_publicMethods_mean}} and Cosine Average 15 - \textit{Figure \ref{fig:agglo_algo_cosine_average_15}}
    \begin{itemize}
        \item There are distinct distributions between the left and right clusters which are reflected in both the footprints. Representing the software features of the target software \textbf{$s\in S$} in the new instance space, when the software features fall in the region of 0 > z\_2 > -0.1 and z\_1 > 0.4, publicMethods\_mean is the most important feature in determining whether Cosine Average 15 is the \newer{most suitable} clustering algorithm.
    \end{itemize}
    \item AnonymousClassesQty\_mean - \textit{Figure \ref{fig:agglo_feature_anonymousClassesQty_mean}} and Cosine Average 7 - \textit{Figure \ref{fig:agglo_algo_cosine_average_7}}
    \begin{itemize}
        \item There are distinct distributions between the left and right clusters which are reflected in both the footprints. Representing the software features of the target software \textbf{$s\in S$} in the new instance space, when the software features fall in the region of z\_2 < -0.6 and z\_1 > 0.1, AnonymousClassesQty\_mean is the most important feature in determining whether Cosine Average 7 is the \newer{most suitable} clustering algorithm.
    \end{itemize}
\end{itemize}

However, there are certain features such as Static Methods\_std - \textit{Figure \ref{fig:agglo_feature_staticMethods_std}} and SubClassesQty\_mean - \textit{Figure \ref{fig:agglo_feature_anonymousClassesQty_mean}} that do not have a clear distinct distribution among the clusters. These features do not have any similar distribution patterns when compared to the prioritised agglomerative algorithms as well.

\subsubsection{Findings from Agglomerative Footprints}
Softwares with a higher value of staticMethods, publicMethods, privateMethods, and modifiers are more complex which leads to more opportunities for remodularisation~\cite{chidamber1994}. While there is no existing literature that identifies any correlation between the number of methods and the performance of software clustering algorithms, the work by \cite{scanniello2013class,scanniello2010using,chidamber1994} discusses how metrics related to the size of the software can be effectively used to measure the quality of object-oriented software systems and for fault prediction. Our approach provides clear evidence of the impact these software features have on the effectiveness of software clustering and remodularisation techniques. 

For agglomerative clustering, the algorithm carries out clustering based on the provided distance matrix given by \new{the \textit{Depends} tool}, where classes with high functional dependency (i.e., minimal distance) would be clustered together to form a cluster (subsystem). The \textit{Depends} tool measures dependencies between classes by analysing method invocation, type casting, and variable containment, which are strongly correlated with the seven main software features that we have identified - staticMethods, privateMethods, subClassesQty, CBO (Coupling Between Objects), modifiers, publicMethods, and anonymousClassQty. When there are more methods and assignment operations in a class, the probability for the method operations and variable assignments to involve instances from another class is higher. With the richer dependency information extracted from \textit{Depends}, we can better illustrate the interrelationships between classes in the analysed software. 


Based on the results, we are able to observe very distinct clusters from the footprints for the prioritised algorithms. When we investigate the distribution of software features in Figures~\ref{fig:agglo_feature_modifiers_max} and~\ref{fig:agglo_feature_publicMethods_mean}, we see a clear gradient of change from top left to bottom right and from bottom left to top right respectively. This means that we are able to clearly assign the \newer{most suitable} algorithms to these projects based on these software features - modifier\_max and publicMethods\_mean respectively. 

Intuitively, this makes sense, as classes with a high number of public methods and modifiers have more opportunities to be remodularised, by splitting large and complex classes into smaller ones with less methods. 

SubClassesQty and Coupling Between Objects, on the other hand, increase the complexity of the software remodularisation problem. Intuitively, this makes sense, as classes with the presence of SubClasses and high coupling (CBO) makes it harder for us to separate this classes during the clustering process. Hence, the findings from this section help provide answers for the first research question. 

\begin{table}[h!]
\centering
\caption{Performance of agglomerative linkage distribution.}
\begin{tabular}{ |c|c| } 
 \hline
 \textbf{Algorithms} & \textbf{Number of times algorithm is prioritised} \\ 
 \hline
 Average & 48 \\
 \hline
 Complete & 52 \\ 
 \hline
 Single & 200 \\ 
 \hline
\end{tabular}
\label{table:agglo_linkage_distribution}
\end{table}

As for the linkage algorithm, our results show that single linkage triumphs over average and complete linkage in more than 50\% of the projects, reflected in Table \ref{table:agglo_linkage_distribution}. However, single linkage is not prioritised as compared to average linkage based on the SVM model, which is due to the overall SVM accuracy, precision, and recall of each individual algorithm as discussed above. Single linkage algorithm belongs to the $c_3$ cluster (high accuracy, high precision, and low recall), \new{where the SVM model generally prioritises} other algorithms due to the low recall of single linkage algorithms. This is interesting because the single linkage algorithm tends to form large and less coupled clusters. Upon further investigation, we found that the ground truth extracted from the analysed projects tends to have a large directory structure as well, which contributes toward the finding. 

While the work by Maqbool \cite{maqbool2007hierarchical} claimed that the complete linkage algorithm is capable of forming the best software clustering results in terms of cluster cohesiveness, they used binary clustering features (identify the presence or absence of similar features) to identify the interrelationships between software entities. On the other hand, our work utilises quantifiable measures to assign a relative weight to indicate the strength of dependencies between classes. Apart from that, when running the same linkage algorithm on 10 previous releases of the examined software, we found that single linkage produces much more stable results as compared to complete linkage and average linkage algorithms, which largely agrees with the observation found in the work by \cite{tzerpos2000stability}. Since single linkage outperforms other linkage algorithms in most of the scenarios, \new{we do not include the illustration} of the other footprint visualisation in this paper. 

\new{The work by Tzerpos et al. \cite{tzerpos2000stability} stated that single linkage forms the least cohesive cluster. However, we would like to argue that the experiments conducted by the authors are performed on software written in C programming languages. Hence, the same might not be applicable to modern Java-based systems that possess a completely different structure compared to software written in C. We theorise that the feature extraction tool used to capture the relationship between the clustering entities (classes) enables us to extract richer information on the interaction between the classes, thus producing slightly different results as compared to the work by Tzerpos et al.}

\new{To provide a simple illustration on how E-SC4R can effectively recommend the suitable clustering algorithm from the pool of choices, we have compared the E-SC4R framework against some of the baseline agglomerative hierarchical clustering algorithms. Table \ref{table:baseline_comparison} shows a comparison between the MoJo values for "Average Euclidean 10", "Complete Cosine 10", "Single Manhattan 10", and the configuration recommended by E-SC4R. The lower the MoJo value, the more suited the algorithm configuration is for the specific project. The last column in Table \ref{table:baseline_comparison} is the configuration suggested by E-SC4R for the target software. We will like to note that due to space constraints, we are unable to show all the configurations compared against the ones recommended by E-SC4R. The complete information on all clustering results are available on our GitHub page.

By using the proposed framework, developers or researchers can easily identify \newer{a suitable} clustering algorithm and its configuration instead of adopting an exhaustive or trial-and-error approach which is tedious and error prone. }

\begin{table*}
    \caption{\new{Baseline comparison.}}
    \centering
    \begin{tabular}{|c|c|c|c|c|c|c|}
        \hline
    	Project Name & \shortstack{MoJo for \\Average \\Euclidean 10} & \shortstack{MoJo for\\ Complete\\ Cosine 10} & \shortstack{MoJo for \\Single\\ Manhattan 10} & \shortstack{MoJo for E-SC4R\\Recommended \\Configuration} & \shortstack{E-SC4R \\Recommended Configuration} \\
    	\hline
    	bkromhout-realm-java & 88 & 129 & 52 & 43 & manhattan average 25 \\
    	\hline
    	btraceio-btrace & 240 & 301 & 75 & 53 & manhattan average 25 \\
    	\hline
    	bytedeco-javacpp & 41 & 75 & 14 & 8 & manhattan average 25 \\
    	\hline
    	codecentric-spring-boot-admin & 35 & 65 & 21 & 16 & manhattan average 25 \\
    	\hline
    	codenvy-legacy-che-plugins & 1933 & 1358 & 414 & 331 & manhattan average 25 \\
    	\hline
    	coobird-thumbnailator & 42 & 51 & 27 & 21 & manhattan average 25 \\
    	\hline
    	dropwizard & 672 & 429 & 172 & 121 & manhattan average 25\\
    	\hline
    	dropwizard-metrics & 98 & 145 & 58 & 37 & manhattan average 25 \\
    	\hline
    	evant-gradle-retrolambda & 3 & 3 & 3 & 3 & cosine average 7 \\
    	\hline
    	facebook-android-sdk & 173 & 210 & 194 & 120 & cosine average 15 \\
    	\hline
    	facebook-java-business-sdk & 137 & 185 & 170 & 137 & manhattan average 25 \\
    	\hline
    	facebook-fresco & 307 & 388 & 168 & 140 & manhattan average 25 \\
    	\hline
    	facebook-litho & 1058 & 990 & 250 & 164 & manhattan average 25 \\
    	\hline
    	facebook-react-native-fbsdk & 1 & 1 & 1 & 1 & manhattan average 25 \\
    	\hline
    	google-cdep & 56 & 115 & 30 & 19 & manhattan average 25 \\
    	\hline
    	google-dagger & 91 & 138 & 63 & 47 & manhattan average 25 \\
    	\hline
    	google-error-prone & 639 & 830 & 418 & 370 & manhattan average 25 \\
    	\hline
    	google-gitiles & 48 & 43 & 17 & 8 & manhattan average 25 \\
    	\hline
    	google-openrtb & 19 & 20 & 13 & 8 & manhattan average 25 \\
    	\hline
    	google-openrtb-doubleclick & 8 & 14 & 8 & 6 & manhattan average 25 \\
    	\hline
    	grpc-grpc-java & 183 & 160 & 75 & 55 & manhattan average 25 \\
    	\hline
    	havarunner-havarunner & 20 & 23 & 17 & 13 & manhattan average 25\\
    	\hline
    	immutables-immutables & 300 & 412 & 150 & 126 & manhattan average 25\\
    	\hline
    	ionic-team-capacitor & 18 & 42 & 14 & 12 & manhattan average 25\\
    	\hline
    	jankotek-mapdb & 47 & 89 & 25 & 15 & manhattan average 25 \\
    	\hline
    	javafunk-funk & 19 & 50 & 17 & 14 & cosine average 15 \\
    	\hline
    	javaparser & 657 & 792 & 251 & 216 & manhattan average 25 \\
    	\hline
    	permissions-dispatcher & 7 & 10 & 8 & 7 & manhattan average 25 \\
    	\hline
    	pxb1988-dex2jar & 37 & 63 & 30 & 26 & manhattan average 25 \\
    	\hline
    	web3j & 280 & 378 & 90 & 60 & manhattan average 25 \\
    	\hline
    \end{tabular}
    \label{table:baseline_comparison}
\end{table*}

\new{
\begin{tcolorbox}
\textbf{Summary of Agglomerative Footprints Visualisation:} 
Agglomerative clustering algorithms are most impacted by staticMethods, privateMethods, subClassesQty, cbo, modifiers, publicMethods and anonymousClassQty. Single linkage outperforms average and complete linkage.
\end{tcolorbox}
}

\subsection{Bunch Footprints Visualisation}
\begin{figure*}[!ht]
    \centering
    \includegraphics[width=0.55\linewidth]{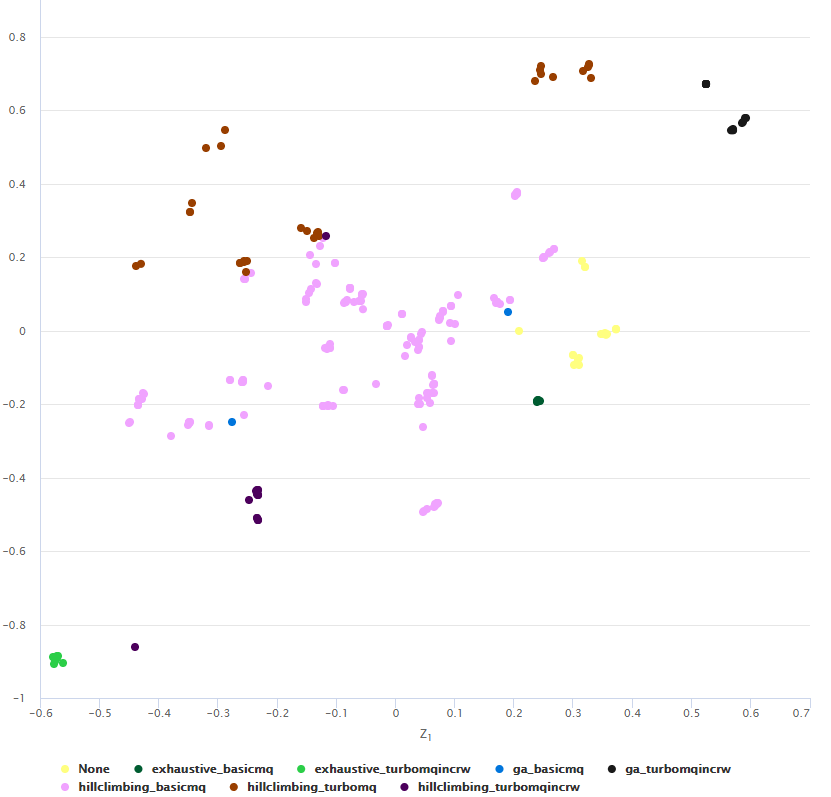}
    \caption{Bunch algorithm selection.}
    \label{fig:bunch_svm_selection}
\end{figure*}

Figure \ref{fig:bunch_svm_selection} illustrates the footprint generated from Bunch algorithm.

The coordinate system that defines the new instance space is defined as:

\begin{equation}
\small
\label{eq:bunch_pca_equation}
\begin{bmatrix}
	z_1 \\
	z_2
\end{bmatrix}
= 
 \begin{bmatrix}
	{0.1127} & {0.0705} \\
	{0.2152} & {0.049} \\
	{-0.0411} & {0.2699} \\
\end{bmatrix}
 \begin{bmatrix}
	{\text{RFC mean}}  \\
	{\text{staticMethods mean}} \\
	{\text{stringLiteralsQty mean}}
\end{bmatrix}
\end{equation}

The three software features in Equation~\ref{eq:bunch_pca_equation} are identified as the most impactful on the performance of the Bunch clustering algorithms used for software remodularisation. staticMethods and stringLiteralsQty are size related metrics, while RFC is a coupling and complexity related metric. All three metrics are correlated because larger projects tend to have more complex classes with higher number of methods that leads to higher RFC.

An interesting finding in Figure~\ref{fig:bunch_svm_selection} is the cluster of projects (yellow color) where none of the Bunch clustering algorithm is predicted to perform well, labelled as "None". The footprint of this cluster is in the "grey zone", where all three software features -- RFC, staticMethods, and stringLiteralsQty, have medium scores such that $z_2$ is in the range between 0.2 < $z_1$ < 0.4 and -0.2 < $z_2$ < 0.2. This is an indication of the specialisation of the Bunch clustering algorithm, and provides evidence that these methods are good at solving extreme cases. This finding highlights another important aspect of our methodology; by analysing the strengths and weaknesses of existing software clustering techniques, we could identify areas that require improvement. In this case, it is evident that there is a gap in software clustering techniques which are able to solve problems that have a medium number of RFC, staticMethods, and stringLiteralsQty. 

\subsubsection{Bunch PCA Relationship between Features and Clusters}
\begin{figure*}[!ht]
\centering
 \subfigure[Footprints of Exhaustive BasicMQ.]{\includegraphics[width=0.47\linewidth]{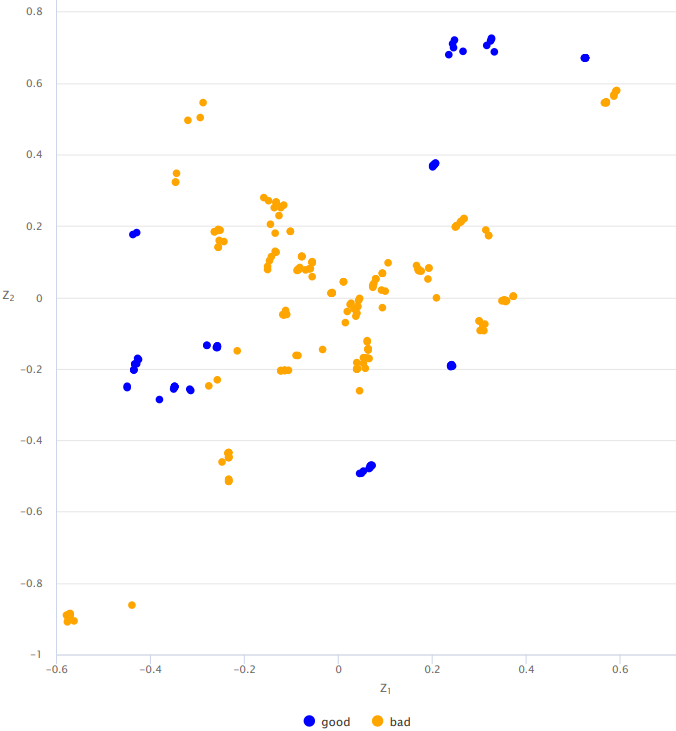}\label{fig:bunch_algo_exhaustive_basicmq}}
 \subfigure[Footprints of Exhaustive TurboMQIncrW.]{\includegraphics[width=0.47\linewidth]{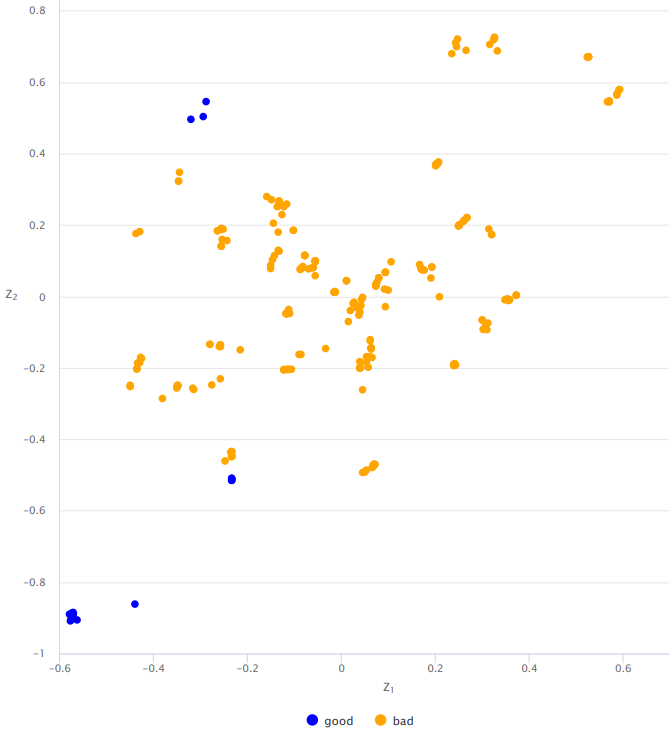}\label{fig:bunch_algo_exhaustive_turbomqincrw}}
 \subfigure[Footprints of GA BasicMQ.]{\includegraphics[width=0.47\linewidth]{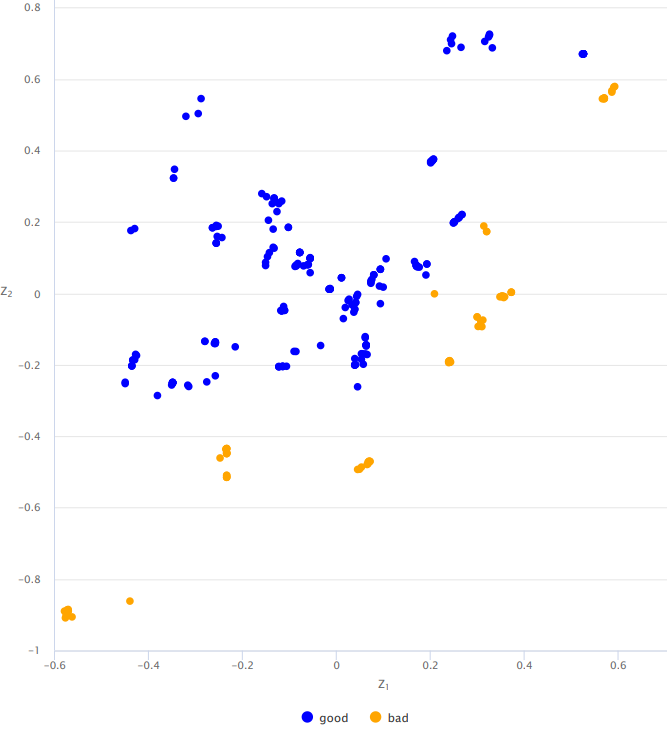}\label{fig:bunch_algo_ga_basicmq}}
 \subfigure[Footprints of GA TurboMQIncrW.]{\includegraphics[width=0.47\linewidth]{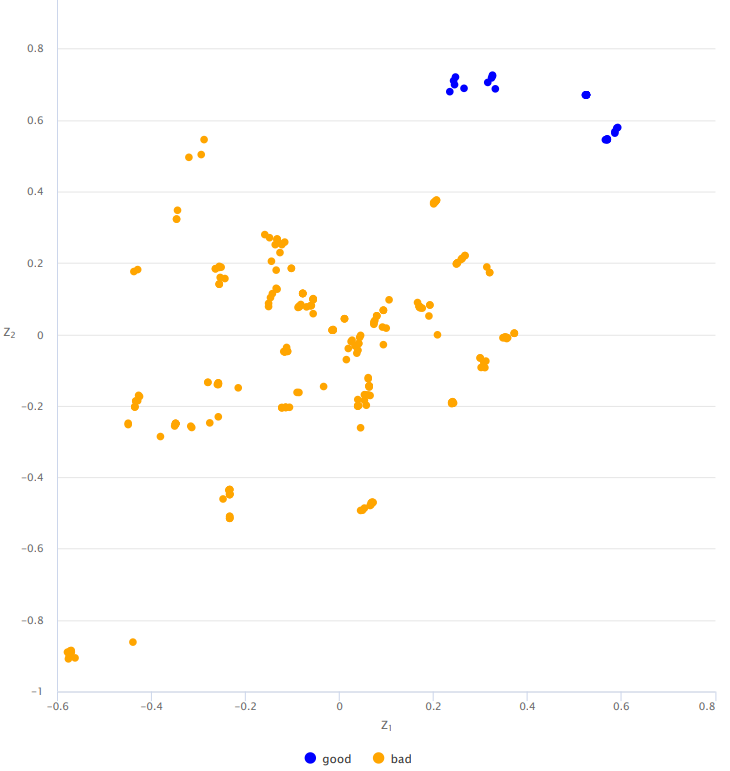}\label{fig:bunch_algo_ga_turbomqincrw}}
 \caption{Bunch algorithm footprint visualisation - Part 1.\label{fig:footprint_bunch_algo_a}}
 \end{figure*}

\begin{figure*}[!ht]
\centering
 \subfigure[Footprints of HillClimbing BasicMQ.]{\includegraphics[width=0.47\linewidth]{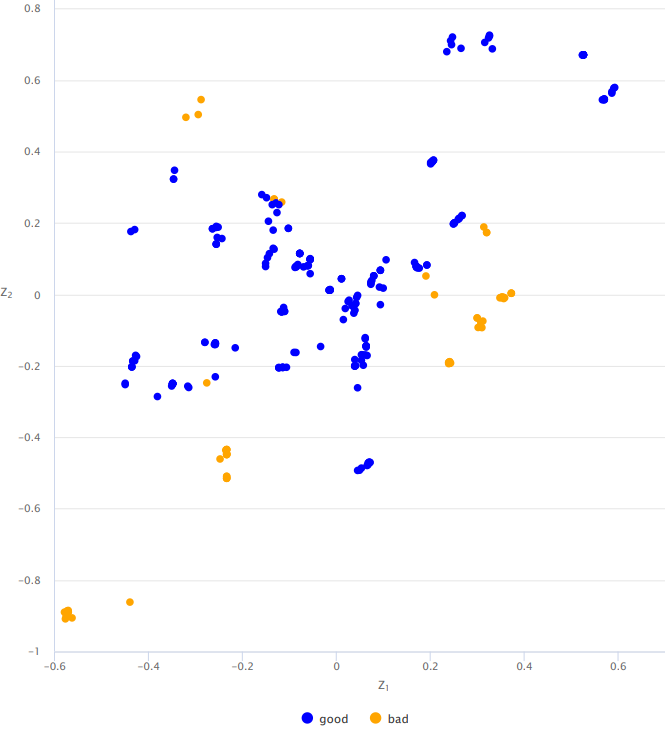}\label{fig:bunch_algo_hillclimbing_basicmq}}
 \subfigure[Footprints of HillClimbing TurboMQ.]{\includegraphics[width=0.47\linewidth]{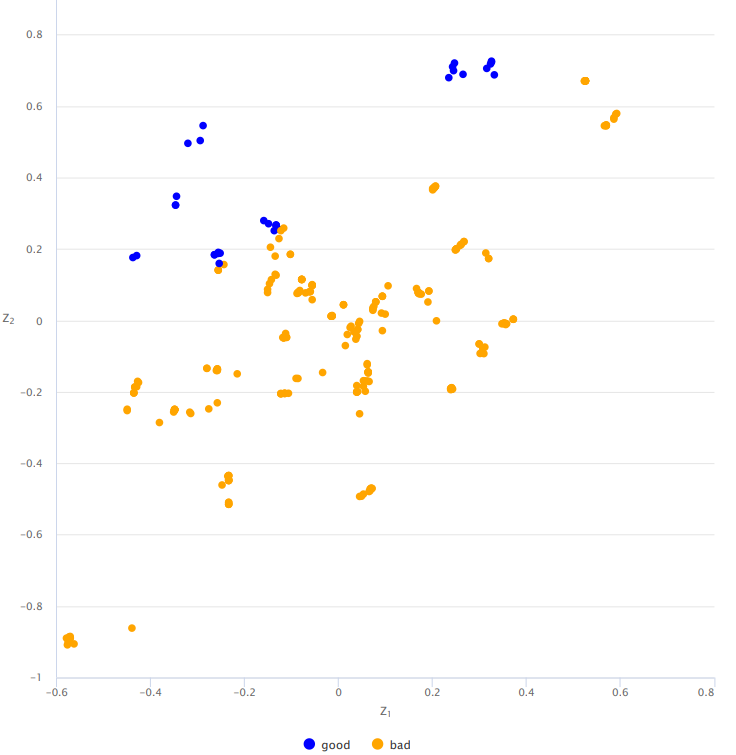}\label{fig:bunch_algo_hillclimbing_turbomq}}
 \subfigure[Footprints of HillClimbing TurboMQIncrW]{\includegraphics[width=0.47\linewidth]{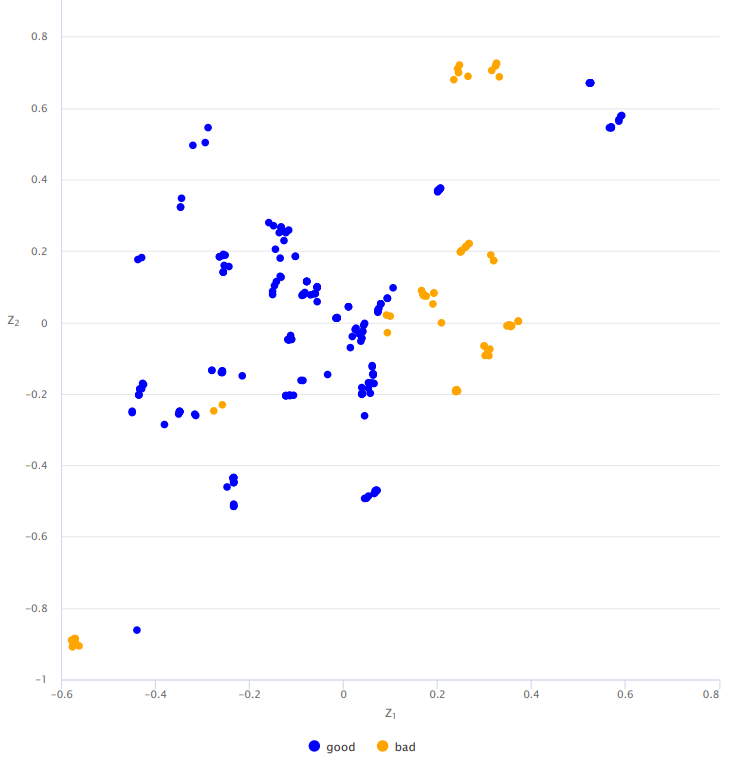}\label{fig:bunch_algo_hillclimbing_turbomqincrw}}
\caption{Bunch algorithm footprint visualisation - Part 2.\label{fig:footprint_bunch_algo_b}}
\end{figure*}

\begin{figure*}[!ht]
 \centering
 \subfigure[Distribution of RFC\_mean.]{\includegraphics[width=0.47\linewidth]{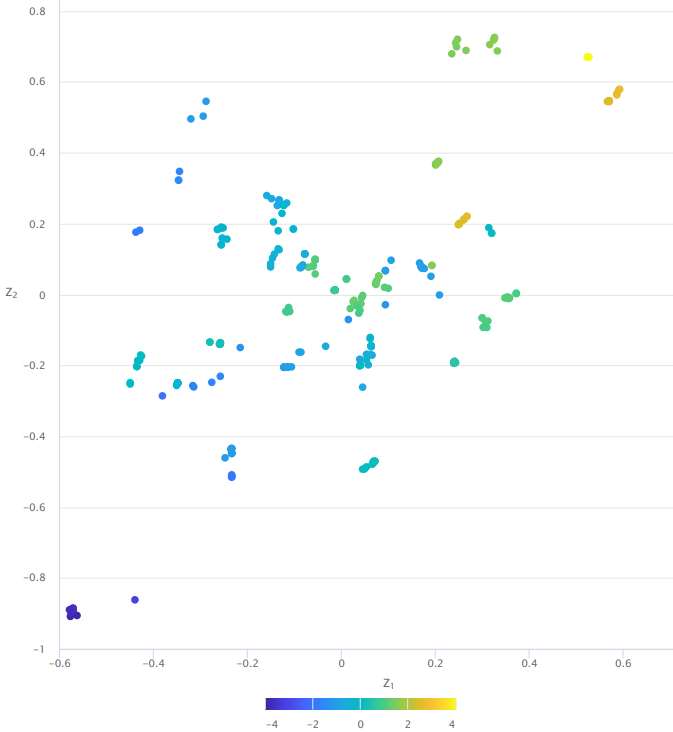}\label{fig:bunch_feature_rfc_mean}}
 \subfigure[Distribution of StaticMethods\_mean.]{\includegraphics[width=0.47\linewidth]{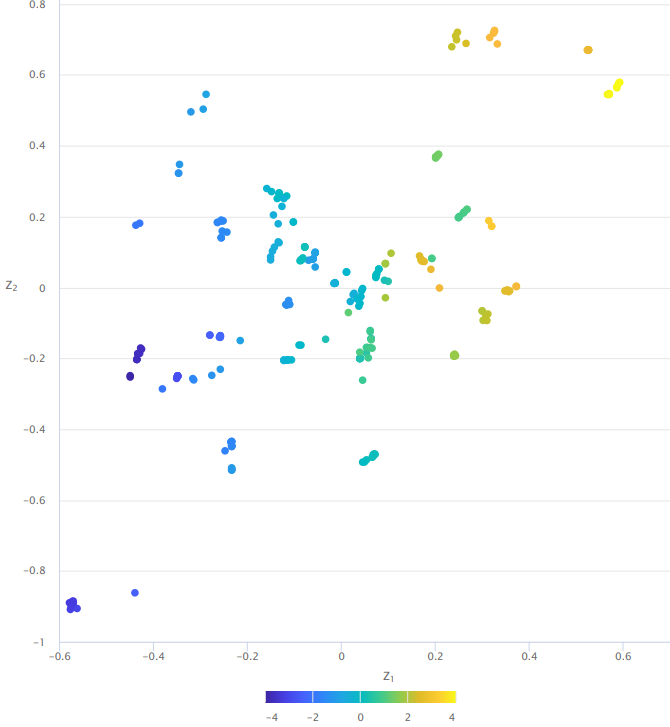}\label{fig:bunch_feature_staticMethods_mean}}
 \subfigure[Distribution of StringLiteralsQty\_mean.]{\includegraphics[width=0.47\linewidth]{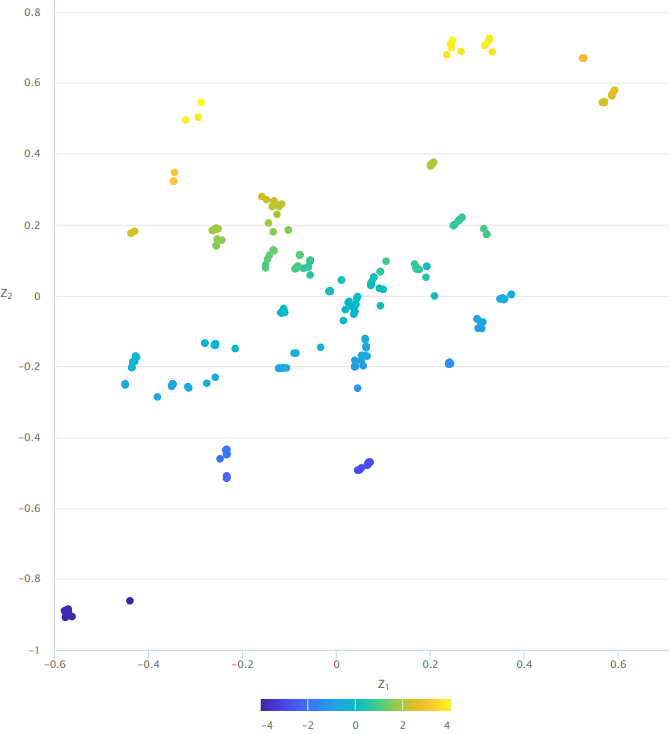}\label{fig:bunch_feature_stringLiteralsQty_mean}}\\
 \caption{Bunch software features footprint visualisation.\label{fig:footprint_bunch_features}}
 \end{figure*}

Using the new coordinate system for Bunch, we visualise the footprints of the different techniques as shown in Figures~\ref{fig:footprint_bunch_algo_a}, \ref{fig:footprint_bunch_algo_b} and \ref{fig:footprint_bunch_features}. To better understand why certain clustering algorithms or parameters works better for software projects in the cluster, we did a side-by-side comparison of feature and SVM model performance to try and draw correlations.

By comparing Figures \ref{fig:footprint_agglo_algo}, \ref{fig:footprint_agglo_features_a} and \ref{fig:footprint_agglo_features_b} we are able to draw similarities between the algorithm's footprint patterns and the distribution of the values of the features. This means that these features are the most important when determining priority of the algorithms.
\begin{itemize}
    \item StringLiteralsQty\_mean - \textit{Figure \ref{fig:bunch_feature_stringLiteralsQty_mean}} and HillClimbing TurboMQ - \textit{Figure \ref{fig:bunch_algo_hillclimbing_turbomq}}
    \begin{itemize}
        \item There are distinct distributions between the top left and bottom right clusters which are reflected in both the footprints. Representing the software features of the target software \textbf{$s\in S$} in the new instance space, when the software features fall in the region of z\_2 > 0.4 and z\_1 < 0.4, StringLiteralsQty\_mean is the most important feature in determining whether HillClimbing TurboMQ is the \newer{most suitable} clustering algorithm.
    \end{itemize}
    \item RFC\_mean - \textit{Figure \ref{fig:bunch_feature_rfc_mean}} and GA TurboMQIncrW - \textit{Figure \ref{fig:bunch_algo_ga_turbomqincrw}}
    \begin{itemize}
        \item There are distinct distributions between the right and left clusters which are reflected in both the footprints. Representing the software features of the target software \textbf{$s\in S$} in the new instance space, when the software features fall in the region of z\_2 > 0.5 and z\_1 > 0.4, RFC\_mean is the most important feature in determining whether GA TurboMQIncrW is the \newer{most suitable} clustering algorithm.
    \end{itemize}
    \item StaticMethods\_mean - \textit{Figure \ref{fig:bunch_feature_staticMethods_mean}} and GA BasicMQ - \textit{Figure \ref{fig:bunch_algo_ga_basicmq}}
    \begin{itemize}
        \item There are distinct distributions between the bottom right and top left clusters which are reflected in both the footprints. Representing the software features of the target software \textbf{$s\in S$} in the new instance space, when the software features fall in the region of z\_2 > 0.4 and z\_1 > 0.3, StaticMethods\_mean is the most important feature in determining whether GA BasicMQ is the \newer{most suitable} clustering algorithm.
    \end{itemize}
\end{itemize}

\subsubsection{Findings from Bunch Footprints Visualisation}
For Bunch, the algorithm carries out clustering based on source code analysis. Bunch uses a family of source code analysis tools (supports C, C++ and Java) that is based on an entity relationship model, where the source code is scanned and a relational database is constructed to store the entities and relations \cite{mitchell2006}. Bunch also assumes that all relation types have an equal weight. Hence, when taking into account variable references or global variables, the amount of StringType literals can be a distinguishable feature. This is why StringLiteralsQty turns out to be an important feature within Bunch where the more String variables that are present within the software, there will be more relationships that the source code analysis tool can identify, \new{which subsequently helps to correctly identify the distribution of the correct clusters}. At the same time, features such as staticMethods and RFC which contributes to the richness of information within the entity relationship model is also important because they can better illustrate the interrelationships between classes.

By looking at the footprint visualisations for the prioritised Bunch algorithm, our results agree to a certain extent with the authors' claim that exhaustive clustering algorithm only works well for small systems \cite{mitchell2002heuristic},
and the hill-climbing algorithm performs well for most software systems \cite{mitchell2006}. The footprints of Exhaustive BasicMQ (Figure \ref{fig:bunch_algo_exhaustive_basicmq}) and Exhaustive TurboMQIncrW (Figure \ref{fig:bunch_algo_exhaustive_turbomqincrw}) shows that there are significantly larger clusters of instances where it is labelled as "Good" by the SVM model when the values for $z\_1$ and $z\_2$ are small (low RFC, staticMethods, and stringLiteralsQty, shown on the bottom left corner of Figure \ref{fig:footprint_bunch_algo_a}). This shows that exhaustive clustering works better on software projects classified as small based on the software features. One good example is shown on Figure \ref{fig:bunch_algo_exhaustive_turbomqincrw}, where only Exhaustive TurboMQIncrW is able to show "Good" results when $z\_1$ is between -0.4 to -0.6, and $z\_2$ is between -0.8 to -1.0.

On the other hand, when the size-related metrics of staticMethods and stringLiteralsQty are high (top right corner of Figure \ref{fig:bunch_feature_staticMethods_mean} and Figure \ref{fig:bunch_feature_stringLiteralsQty_mean}), the footprints of HillClimbing BasicMQ (Figure \ref{fig:bunch_algo_hillclimbing_basicmq}) and TurboMQIncrW (Figure \ref{fig:bunch_algo_hillclimbing_turbomqincrw}) shows that instances are labelled as "Good". This suggests that HillClimbing algorithms performs well on large sized projects. As such, our E-SC4R framework not only reaffirms that hill climbing approach is well suited for large size projects, we further discover that the value of RFC, staticMethods, and stringLiteralsQty can be a good indicator for researchers to decide whether or not to choose exhaustive Bunch or hill climbing Bunch when performing software remodularisation.

Another interesting finding from the footprint visualisations for the prioritised Bunch algorithms is that the same algorithm but with a different calculator is able to cater to a entirely different software type. For example, GA BasicMQ performs well for most software systems, but is unable to cater to extremely large software systems. GA TurboMQIncrW on the other hand, only performs well on extremely large software systems. Similarly, for HillClimbing, HillClimbing BasicMQ and HillClimbing TurboMQIncrW performs well for most software systems, but HillClimbing TurboMQ speficially caters for software with size metrics that are extremely high in the $z\_1$ range ($z\_1 > 0$) and in the middle of the $z\_2$ range ($-1 < z\_2 < 1$).

\new{
\begin{tcolorbox}
\textbf{Summary of Bunch Footprints Visualisation:} 
Bunch clustering algorithms are most impacted by StaticMethods, StringLiteralQtys, RFC. Exhaustive algorithm works better with small systems. Hill-climbing algorithm performs generally well across all systems.The calculator for the algorithm plays a huge role in determining which software type the configuration of the algorithm is best suited for. 
\end{tcolorbox}
}

\subsection{Summary of Experiment Results}

By analysing the strengths and weaknesses of existing software clustering techniques, we could identify areas in current software remodularisation and architecture recovery research that requires improvement. Based on our experiment results and footprints, it is evident that there is a gap in clustering techniques that \newer{takes software features/metrics as a feature during the clustering process}. For example, from Figure \ref{fig:agglo_feature_cbo_mean}, \newer{based on our selected software systems and experiment design,} existing agglomerative clustering techniques are unable to cluster software projects based on Coupling Between Objects (CBO). We are not suggesting that CBO is not a good indicator for software remodularisation problems, but rather, existing software clustering algorithms do not benefit from CBO as a feature during the clustering process.

The ability to identify \newer{a suitable} \new{algorithm and configuration from our existing pool of choices} to configure software clustering algorithms (such as the linkage algorithm \new{and distance metric}) based on the characteristics of different software can help software developers and maintainers to reduce the time and effort needed to perform software remodularisation in a more effective manner, rather than having to resort to intuition or trial and error, both of which have far lower accuracy rates and and it is either resource intensive when evaluating multiple approaches, or they might simply just work with a sub-optimal cluster set. The experiment results also suggest that on a larger scale (analysis of more projects and more distinct identification of domains or metrics to classify a pool of software), it would be possible to, at the minimum, eliminate approaches or parameters that are already known to be sub-optimal. The visualisation has the potential to show the most optimum approach that a software maintainer can adopt, thus improving the accuracy of clustering results as well as saving resources that would otherwise potentially be wasted. 

\vspace{2mm}
\begin{tcolorbox}[breakable, enhanced jigsaw,rounded corners, parbox=false]
\textbf{RQ1:} The strengths and weaknesses of agglomerative hierarchical and Bunch clustering algorithms can be evaluated from the footprints generated. Drawing examples from the Bunch footprint, the strength of the algorithm can be seen in its ability to cluster software accurately through RFC, staticMethods, and stringLiteralsQty metrics, where identifiable and significant clusters can be found from the feature footprints
The weakness of the algorithm can be instead seen through clusters that are labelled as "None" during the prioritised algorithm footprint visualisation. This shows that the techniques are good at solving extreme cases, but unable to properly cluster software projects with medium-sized metrics, identifying the gap in the aforementioned software clustering techniques. As such, the answers to both research questions are discussed below.

\end{tcolorbox}

\vspace{2mm}
\begin{tcolorbox}[breakable, enhanced jigsaw,rounded corners, parbox=false]
\textbf{RQ2:} When using MoJoFM as the evaluation criteria, agglomerative hierarchical clustering proves to be the clear winner when compared to different variations and configuration of Bunch clustering algorithm. It is when the two algorithms are evaluated separately using E-SC4R and using the conclusions drawn from the footprints, we are able to more objectively select the \newer{most suitable} clustering technique based on the software features of each test subject. Based on our results, modifier\_max, publicMethods\_mean, and anonymousClassQty\_mean are the three most prevalent software features that affect the performance of agglomerative hierarchical clustering algorithm, while RFC\_mean, staticMethods\_mean, and stringLiteralsQty\_mean plays essential role in deciding the type of Bunch algorithm to aid in software remodularisation and architecture recovery. As such, using the identified software features as indicators, they can be used to aid researchers in selecting the \newer{most suitable} clustering technique, which depends on the characteristic of the software remodularisation and architecture recovery problem. 
\end{tcolorbox}

\section{Threats to Validity}
Based on the classification schema of Runeson et al. \cite{runeson2012case}, \textbf{Construct Validity} in our case refers to whether all relevant parameters for hierarchical clustering and Bunch clustering algorithm have been explored to visualise the footprint for software remodularisation. To mitigate this risk, we considered a plethora of parameters such as number of projects, number of revisions, different linkage algorithms, distance metrics, and search-based fitness function. Besides, we take into consideration the past releases of the examined software in generating the ground truth.\newer{
With regards to the selection of software features, we choose to use direct software features such as CK suite of metrics (mainly code-related metrics) to represent the characteristics of the analysed software because it has been proven in existing studies that these features are able to reliably present the characteristics of software with respect to the effectiveness of clustering algorithms \cite{naseem2019euclidean, maqbool2007hierarchical}. Indirect software features such as the number of bugs, number of commits, and number of active contributors (mainly project-related metrics) are not used in this study due to a lack of literature that indicates a strong relationship between such features with the effectiveness of software clustering algorithms.
}

\textbf{Internal Validity} is related to the examination of causal relations. Our results pinpoint the particular software features that affect the effectiveness of the agglomerative and Bunch clustering method of the analysed project, which is not inferred from causal relationships. 

With respect to \textbf{External Validity}, the risk is mitigated by selecting a pool of projects that are well-known and popular in the the open-source community (project selected based on the number of stars on GitHub) forming a representative sample for analysis. \new{In order to provide more information about the quality of the chosen project, \textit{Sonarqube} \cite{campbell2013sonarqube} has been used to analyse the quality of the chosen projects in terms of the number of bugs, code smells, and code duplication presented in Table~\ref{table:project_sqale}. This is to demonstrate the application on E-SC4R on a variety of software projects in terms of size and quality.} However, a replication of this study in a larger scale that comprises projects written in different languages would be valuable in verifying the current findings. We have created a replication package on our GitHub page. 

\begin{table*}
    \caption{\new{Quality metrics of the chosen projects extracted from SonarQube.}}
    \centering
    \begin{tabular}{|c|c|c|c|c|c|c|c|c|}
        \hline
    	Project & Bugs & \shortstack{Bugs\\Rating} & \shortstack{Code Smells\\(thousands)} & \shortstack{Code Smells\\ Rating} & \shortstack{Duplications\\(\%)} & \shortstack{Duplications\\Rating} & \shortstack{Lines\\(thousands)} & SizeRating \\
    	\hline
    	\shortstack{bkromhout-realm-\\java} & 137 & E & 4.5 & A & 9.6 & C & 52 & M\\
    	\hline
    	btraceio-btrace & 78 & E & 2.3 & A & 2.5 & A & 46 & M \\
    	\hline
    	bytedeco-javacpp & 301 & E & 1.9 & A & 5.9 & C & 27 & M \\
    	\hline
    	\shortstack{codecentric-spring-\\boot-admin} & 152 & E & 1.2 & A & 1.8 & A & 20 & M\\
    	\hline
    	\shortstack{codenvy-legacy-\\che-plugins} & 548 & E & 36 & A & 11.1 & D & 477 & L\\
    	\hline
    	\shortstack{coobird-\\thumbnailator} & 70 & E & 1.8 & A & 10.2 & D & 21 & M\\
    	\hline
    	dropwizard & 19 & E & 1.8 & A & 3.2 & B & 61 & M\\
    	\hline
    	dropwizard-metrics & 27 & D & 0.9 & A & 23.4 & E & 36 & M\\
    	\hline
    	\shortstack{evant-gradle-\\retrolambda} & 1 & E & 0.02 & A & 26.8 & E & 2 & S\\
    	\hline
    	\shortstack{facebook-\\android-sdk} & 88 & E & 2.1 & A & 1.6 & A & 73 & M\\
    	\hline
    	\shortstack{facebook-java-\\business-sdk} & 26 & E & 25 & A & 36.8 & E & 381 & L\\
    	\hline
    	facebook-litho & 257 & E & 38 & A & 30.9 & E & 568 & XL\\
    	\hline
    	\shortstack{facebook-react-\\native-fbsdk} & 0 & A & 0.03 & A & 0 & A & 1.5 & S\\
    	\hline
    	google-cdep	 & 49 & E & 1.4 & A & 4.6 & B & 16 & M\\
    	\hline
    	google-dagger & 46 & E & 3.4 & A & 4.1 & B & 104 & L\\
    	\hline
    	google-gitiles & 16 & E & 0.5 & A & 2.9 & A & 16 & M\\
    	\hline
    	google-openrtb & 6 & C & 0.04 & A & 0 & A & 2.3 & S \\
    	\hline
    	\shortstack{google-openrtb-\\doubleclick} & 3 & C & 0.08 & A & 0 & A & 2.8 & S\\
    	\hline
    	grpc-java & 253 & E & 7.8 & A & 7.7 & C & 203 & L\\
    	\hline
    	havarunner & 1 & C & 0.35 & A & 0.9 & A & 3.6 & S\\
    	\hline
    	immutables & 54 & D & 1.7 & A & 2.2 & A & 71 & M\\
    	\hline
    	\shortstack{ionic-team-\\capacitor} & 3 & E & 0.2 & A & 0.4 & A & 6.2 & S\\
    	\hline
    	jankotek-mapdb & 151 & E & 2.3 & A & 6.7 & C & 20 & M\\
    	\hline
    	javafunk-funk & 29 & C & 3 & B & 1.3 & A & 26 & M\\
    	\hline
    	javaparser & 2200 & E & 12 & A & 23.6 & E & 187 & L\\
    	\hline
    	\shortstack{permissions-\\dispatcher} & 1 & C & 0.5 & A & 27 & E & 14 & M\\
    	\hline
    	pxb1988-dex2jar & 34 & E & 2.5 & A & 2.7 & A & 36 & M\\
    	\hline
    	web3j & 90 & E & 2.2 & A & 5.7 & C & 48 & M\\
    	\hline
    \end{tabular}
    \label{table:project_sqale}
\end{table*}

\new{
Ground truth generation plays a vital role in determining the optimum clustering algorithm and configuration from the existing pool. Getting input from domain expert may help reaffirm the validity of our ground truth. However, due to the scope of the project where we experimented on 30 open source projects, it is challenging to get the developers from the open source community to evaluate the ground truth individually for each version for each project. Based on the state-of-the-art, there is no single well-acknowledged method in creating the ground truth for software clustering. One of the most popular approaches, however, is by leveraging on the package structure of the analysed software. Hence, in this research, we have adopted a similar approach to address the problem. We like to note that the proposed E-SC4R framework can work with any kind of clustering algorithm and ground truth, as long as the ground truth is standardised for comparison across the existing pool of clustering algorithms. 

Software systems with only a few directories with a large number of files in each might not validate some of our results as well. Upon examining the existing ground truth that we have, we found that the effect of having a few directories with large number of files is negligible. In fact, the majority of the ground truths that we use for the experiments consist of projects with large numbers of directories (packages), with a small number of files in each of them. We have uploaded some of the ground truth that we used for the experiments on the GitHub page.\footnote{https://github.com/alvintanjianjia/SoftwareRemodularization/tree/master/sample\_groundtruth}

On the other hand, code quality, which includes coding style, readability, level of cohesion, and other indicators are factors that might impact the effectiveness of the presented clustering algorithms. In this research, we have only evaluated the proposed approach on open source systems. While the code quality of open source and real-life industrial systems is very subjective and context dependent (the quality of open source projects can be low when compared to industrial project, and vice versa), we expect that applying the proposed E-SC4R framework on project with different code quality (real-life industrial systems included) will yield different results, mainly because the characteristic of software (i.e. CK metrics of the analysed software) will affect the footprint constructed using our framework. 

The challenge of running the experiments on low code quality projects is the construction of ground truth to validate the clustering algorithm. In our proposed approach and in existing studies, package structure is the most commonly used method to create the artificial ground truth for software clustering. We construct the ground truth by looking at the package structure of the past 10 releases of the software, and find the overlapping and most common directory structure to be used as the ground truth. If the same approach is to be applied to low code quality projects, the ground truth will be heavily skewed and not reliable, which affects the profiling of these systems. With an accurate ground truth generated manually by an expert, the E-SC4R framework is applicable to both open source and industrial projects for reversing the documentation of poorly documented systems with high technical debts.
}

\section{Related Works on Software Clustering for Architecture Recovery}

In general, software clustering consists of the following four steps. First, common clustering features are chosen to determine the similarity between entities (methods, classes, or packages depending on the level of granularity). Second, a similarity measure is chosen to determine the similarity strength between two entities (method invocation, passing of parameters, sharing of variables, etc.) \cite{maqbool2007hierarchical}. Third, a clustering algorithm is chosen to group similar entities together. Finally, a form of validation is required to measure the quality of the clustering results. The results of software clustering can be viewed as a high-level abstraction of the software architecture to aid in software comprehension. All four steps mentioned above play a significant role in determining the quality of clustering results because the selection of different clustering features or metrics will produce substantially different clustering results. 

Although some clustering algorithms produce a single clustering result for any given dataset, a dataset may have more than one natural and optimum clustering result. For instance, source code can only reveal very limited information about the architectural design of a software system since it is a very low-level software artefact. On the other hand, some implementation details might be lost if software packages are being used to represent clustering entities. Hence, identifying the most optimum way to choose the most appropriate clustering algorithm and configure the parameters of the algorithm is a non-trivial task in software remodularisation. 

The work by Deursen and Kuipers \cite{vanDeursen1999} adopted a greedy search method by using mathematical analysis to analyse the structure of cluster entities and identify the clustering features that are shared by them. The proposed approach finds all of the possible combinations of clusters and evaluates the quality of each combination. Agglomerative hierarchical clustering is used in this work. The authors discovered that it is hard to analyse all possible combinations, and useful information might be missing if no attention is given to analyse all the results generated from different dendrogram cutting points. 

In contrast to the greedy search method proposed by Deursen and Kuipers, the work by Fokaefs et al. \cite{Fokaefs} proposed an approach that produces multiple clustering results from which software developers and maintainers can choose the best result based on their experiences. The goal is to decompose large classes by identifying ‘Extract Class’ refactoring opportunities. Extract class is defined as classes that contain many methods without clear functionality. The authors adopted the agglomerative clustering algorithm to generate a dendrogram and cut the dendrogram at several places to form multiple sets of results. The authors argued that clustering algorithms that produce a single result are too rigid and not feasible to fit into the context of software development.

Work by Anquetil and Lethbridge \cite{anquetil1999experiments} attempted to perform agglomerative clustering on source files and found out that using source code alone to aid in software remodularisation yields poor results. In their study, clustering entities are represented in the form of source code. The authors found that the quantity of information, such as the number of variables used in the source code, the dependency between routines, and the data passed and shared by functions helps in improving the reliability of clustering.

The work by Cui and Chae \cite{cui2011applying} attempted to analyse the performance, strengths, and weaknesses of different agglomerative hierarchical clustering algorithms using multiple case studies and setups. The authors conducted a series of experiments using 18 clustering strategies. The clustering strategies are the combination of different similarity measures, linkage methods, and weighting schemes. The case studies comprise 11 systems where source codes were used as the input parameters. The authors found that it is difficult to identify a perfect clustering strategy which can fulfil all the evaluation criteria proposed by the authors.

As discussed in the systematic literature review conducted by Alsarhan et al. \cite{Alsarhan2020}, the selection of clustering algorithms remains a challenging problem in the area of software clustering for remodularisation and architectural recovery.  While there have been attempts to propose guidelines for selecting or rejecting a clustering algorithm for a given software \cite{shtern2009methods}, there is a lack of comprehensive methods for clustering algorithm selection. 

Our work extends existing research in analysing the effectiveness of software clustering techniques by examining what software features impacts the performance of agglomerative and Bunch clustering algorithms. We characterise a software system using software/code features (e.g., depth of inheritance tree, cohesion, and coupling) and determine the most significant features that have an impact on whether a software clustering technique can generate a good clustering result.

\section{Conclusion and Future Work}

Acknowledging the lack of a universal approach in finding the optimum clustering algorithm for any software remodularisation problem given the numerous algorithms that exist in the literature and the various parameters that may be used to configure software clustering algorithms, we are able to provide empirical evidences that help in identifying key characteristics of software/code features that influence the effectiveness of hierarchical and Bunch-based software clustering algorithms.

Given the relatively high cost of running a clustering algorithm on large and complex software systems, the proposed approach optimises the resources spent on software remodularisation. The results in this paper, while promising, are constrained in a number of ways. First, while it is one of the most popular software remodularisation approaches, only agglomerative hierarchical clustering and Bunch clustering methods were assessed in this paper because it is difficult to compare the clustering results produced from different families of clustering algorithms in a fair and unbias manner. \new{The relationships that are extracted from Depends are aggregated in the proposed approach which may lead to the loss of some semantic information. One way to address this problem is by running the experiments multiple times using only one type of the extracted relationship at a time. For example, if we have 14 types of relationships extracted using Depends, we can run it 14 times for each version of the project, and evaluate the effectiveness of different clustering algorithms using each type of relationship (or combination of multiple relationships).}

The creation of the ground truth relies on the past 10 releases of a particular software, which strongly favours legacy software that have more active developers and contributors. Less stable software with radical changes between versions make it difficult to construct a usable ground truth. For future work, a separate method of defining and creating the ground truth can be further explored. \new{An approach similar to the work by Naseem et al. \cite{naseem_deris_maqbool_li_shahzad_shah_2017} by taking a deep dive into 1 or 2 of our generated ground truth with a few senior and experienced developers who have been working on the respective projects would be one of the future directions of this research, to affirm the validity of our ground truth and clustering results.}.

\newer{One of the aims in this research is to analyse the relationships between software features and the effectiveness of software clustering algorithms. In our work, only direct software features which are code-related metrics, were used to represent the characteristics of the analysed software. The inclusion of indirect software features such as the number of bugs, number of commits, number of active contributors, and other project-related metrics in the future would be able to provide us with more insights as to whether these features are good representations of the characteristics of the software system and their effectiveness when used in software clustering algorithms.}

Nonetheless, the experiment results show that our findings are a step forward in the area of software remodularisation to reveal the strengths and weaknesses of different hierarchical clustering and Bunch clustering algorithms, and it is hoped that the work discussed in this paper can serve as a framework for further analysis and improvements to be made.

\section*{Acknowledgement}
This work was carried out within the framework of the research project FRGS/1/2018/ICT01/MUSM/03/1 under the Fundamental Research Grant Scheme provided by the Ministry of Education, Malaysia. 

\bibliographystyle{IEEEtran}
\bibliography{references}

\begin{thebibliography}{10}
\providecommand{\url}[1]{#1}
\csname url@samestyle\endcsname
\providecommand{\newblock}{\relax}
\providecommand{\bibinfo}[2]{#2}
\providecommand{\BIBentrySTDinterwordspacing}{\spaceskip=0pt\relax}
\providecommand{\BIBentryALTinterwordstretchfactor}{4}
\providecommand{\BIBentryALTinterwordspacing}{\spaceskip=\fontdimen2\font plus
\BIBentryALTinterwordstretchfactor\fontdimen3\font minus
  \fontdimen4\font\relax}
\providecommand{\BIBforeignlanguage}[2]{{%
\expandafter\ifx\csname l@#1\endcsname\relax
\typeout{** WARNING: IEEEtran.bst: No hyphenation pattern has been}%
\typeout{** loaded for the language `#1'. Using the pattern for}%
\typeout{** the default language instead.}%
\else
\language=\csname l@#1\endcsname
\fi
#2}}
\providecommand{\BIBdecl}{\relax}
\BIBdecl

\bibitem{hall2018effectively}
M.~Hall, N.~Walkinshaw, and P.~McMinn, ``Effectively incorporating expert
  knowledge in automated software remodularisation,'' \emph{IEEE Transactions
  on Software Engineering}, vol.~44, no.~7, pp. 613--630, 2018.

\bibitem{RN66}
N.~Anquetil and T.~C. Lethbridge, ``Comparative study of clustering algorithms
  and abstract representations for software remodularisation,'' \emph{IEE
  Proceedings - Software}, vol. 150, no.~3, pp. 185--201, 2003.

\bibitem{RN67}
K.~Praditwong, M.~Harman, and Y.~Xin, ``Software module clustering as a
  multi-objective search problem,'' \emph{IEEE Transactions on Software
  Engineering}, vol.~37, no.~2, pp. 264--282, 2011.

\bibitem{wu2005comparison}
J.~Wu, A.~E. Hassan, and R.~C. Holt, ``Comparison of clustering algorithms in
  the context of software evolution,'' in \emph{21st IEEE International
  Conference on Software Maintenance (ICSM'05)}.\hskip 1em plus 0.5em minus
  0.4em\relax IEEE, 2005, pp. 525--535.

\bibitem{RN69}
M.~Fokaefs, N.~Tsantalis, A.~Chatzigeorgiou, and J.~Sander, ``Decomposing
  object-oriented class modules using an agglomerative clustering technique,''
  \emph{IEEE International Conference on Software Maintenance}, pp. 93--101,
  2009.

\bibitem{RN70}
\BIBentryALTinterwordspacing
R.~Aull-Hyde, S.~Erdogan, and J.~M. Duke, ``An experiment on the consistency of
  aggregated comparison matrices in ahp,'' \emph{European Journal of
  Operational Research}, vol. 171, no.~1, pp. 290--295, 2006. [Online].
  Available: \url{httpwww.sciencedirect.comsciencearticlepiiS0377221704005971}
\BIBentrySTDinterwordspacing

\bibitem{andritsos2005information}
P.~Andritsos and V.~Tzerpos, ``Information-theoretic software clustering,''
  \emph{IEEE Transactions on Software Engineering}, vol.~31, no.~2, pp.
  150--165, 2005.

\bibitem{teymourian2020fast}
N.~Teymourian, H.~Izadkhah, and A.~Isazadeh, ``A fast clustering algorithm for
  modularization of large-scale software systems,'' \emph{IEEE Transactions on
  Software Engineering}, 2020.

\bibitem{chong2013efficient}
C.~Y. Chong, S.~P. Lee, and T.~C. Ling, ``Efficient software clustering
  technique using an adaptive and preventive dendrogram cutting approach,''
  \emph{Information and Software Technology}, vol.~55, no.~11, pp. 1994--2012,
  2013.

\bibitem{aghdasifam2020new}
M.~Aghdasifam, H.~Izadkhah, and A.~Isazadeh, ``A new metaheuristic-based
  hierarchical clustering algorithm for software modularization,''
  \emph{Complexity}, vol. 2020, 2020.

\bibitem{mitchell2002heuristic}
B.~S. Mitchell and S.~Mancoridis, \emph{A heuristic search approach to solving
  the software clustering problem}.\hskip 1em plus 0.5em minus 0.4em\relax
  Drexel University Philadelphia, PA, USA, 2002.

\bibitem{mitchell2006}
B.~Mitchell and S.~Mancoridis, ``On the automatic modularization of software
  systems using the bunch tool,'' \emph{IEEE Transactions on Software
  Engineering}, vol.~32, no.~3, pp. 193--208, March 2006.

\bibitem{prajapati2020harmony}
A.~Prajapati and Z.~W. Geem, ``Harmony search-based approach for
  multi-objective software architecture reconstruction,'' \emph{Mathematics},
  vol.~8, no.~11, p. 1906, 2020.

\bibitem{maqbool2007hierarchical}
O.~Maqbool and H.~Babri, ``Hierarchical clustering for software architecture
  recovery,'' \emph{IEEE Transactions on Software Engineering}, vol.~33,
  no.~11, 2007.

\bibitem{shtern2010comparability}
M.~Shtern and V.~Tzerpos, ``On the comparability of software clustering
  algorithms,'' in \emph{Program Comprehension (ICPC), 2010 IEEE 18th
  International Conference on}.\hskip 1em plus 0.5em minus 0.4em\relax IEEE,
  2010, pp. 64--67.

\bibitem{chong2017automatic}
C.~Y. Chong and S.~P. Lee, ``Automatic clustering constraints derivation from
  object-oriented software using weighted complex network with graph theory
  analysis,'' \emph{Journal of Systems and Software}, vol. 133, pp. 28--53,
  2017.

\bibitem{shtern2012}
M.~Shtern and V.~Tzerpos, ``Clustering methodologies for software
  engineering,'' \emph{Advances in Software Engineering}, vol. 2012, p.~1,
  2012.

\bibitem{Oliveira2019}
\BIBentryALTinterwordspacing
C.~Oliveira, A.~Aleti, Y.-F. Li, and M.~Abdelrazek, ``Footprints of fitness
  functions in search-based software testing,'' in \emph{Proceedings of the
  Genetic and Evolutionary Computation Conference}, ser. GECCO '19.\hskip 1em
  plus 0.5em minus 0.4em\relax New York, NY, USA: ACM, 2019, pp. 1399--1407.
  [Online]. Available: \url{httpdoi.acm.org10.11453321707.3321880}
\BIBentrySTDinterwordspacing

\bibitem{oliveira2018mapping}
C.~Oliveira, A.~Aleti, L.~Grunske, and K.~Smith-Miles, ``Mapping the
  effectiveness of automated test suite generation techniques,'' \emph{IEEE
  Transactions on Reliability}, vol.~67, no.~3, pp. 771--785, 2018.

\bibitem{factbase}
M.~Shtern and V.~Tzerpos, ``Factbase and decomposition generation,'' in
  \emph{2011 15th European Conference on Software Maintenance and
  Reengineering}, March 2011, pp. 111--120.

\bibitem{mancoridis1998}
S.~Mancoridis, B.~S. Mitchell, C.~Rorres, Y.~Chen, and E.~R. Gansner, ``Using
  automatic clustering to produce high-level system organizations of source
  code,'' in \emph{Program Comprehension, 1998. IWPC'98. Proceedings., 6th
  International Workshop on}.\hskip 1em plus 0.5em minus 0.4em\relax IEEE,
  1998, pp. 45--52.

\bibitem{harman2005}
M.~Harman, S.~Swift, and K.~Mahdavi, ``An empirical study of the robustness of
  two module clustering fitness functions,'' in \emph{Proceedings of the 7th
  annual conference on Genetic and evolutionary computation}.\hskip 1em plus
  0.5em minus 0.4em\relax ACM, 2005, pp. 1029--1036.

\bibitem{beck2013impact}
F.~Beck and S.~Diehl, ``On the impact of software evolution on software
  clustering,'' \emph{Empirical Software Engineering}, vol.~18, no.~5, pp.
  970--1004, 2013.

\bibitem{schutze2008introduction}
H.~Sch{u}tze, C.~D. Manning, and P.~Raghavan, \emph{Introduction to information
  retrieval}.\hskip 1em plus 0.5em minus 0.4em\relax Cambridge University Press
  Cambridge, 2008, vol.~39.

\bibitem{dhillon2002enhanced}
I.~S. Dhillon, S.~Mallela, and R.~Kumar, ``Enhanced word clustering for
  hierarchical text classification,'' in \emph{Proceedings of the eighth ACM
  SIGKDD international conference on Knowledge discovery and data
  mining}.\hskip 1em plus 0.5em minus 0.4em\relax ACM, 2002, pp. 191--200.

\bibitem{wiggerts1997using}
T.~A. Wiggerts, ``Using clustering algorithms in legacy systems
  remodularization,'' in \emph{Reverse Engineering, 1997. Proceedings of the
  Fourth Working Conference on}.\hskip 1em plus 0.5em minus 0.4em\relax IEEE,
  1997, pp. 33--43.

\bibitem{beck2016identifying}
F.~Beck, J.~Melcher, and D.~Weiskopf, ``Identifying modularization patterns by
  visual comparison of multiple hierarchies,'' in \emph{Program Comprehension
  (ICPC), 2016 IEEE 24th International Conference on}.\hskip 1em plus 0.5em
  minus 0.4em\relax IEEE, 2016, pp. 1--10.

\bibitem{naseem2019euclidean}
R.~Naseem, M.~M. Deris, O.~Maqbool, and S.~Shahzad, ``Euclidean space based
  hierarchical clusterers combinations an application to software clustering,''
  \emph{Cluster Computing}, vol.~22, no.~3, pp. 7287--7311, 2019.

\bibitem{Alsarhan2020}
Q.~Alsarhan, B.~S. Ahmed, M.~Bures, and K.~Z. Zamli, ``Software module
  clustering an in-depth literature analysis,'' \emph{IEEE Transactions on
  Software Engineering}, pp. 1--1, 2020.

\bibitem{wen2003optimal}
Z.~Wen and V.~Tzerpos, ``An optimal algorithm for mojo distance,'' in
  \emph{Program Comprehension, 2003. 11th IEEE International Workshop
  on}.\hskip 1em plus 0.5em minus 0.4em\relax IEEE, 2003, pp. 227--235.

\bibitem{CKmetric}
M.~Aniche, \emph{Java code metrics calculator (CK)}, 2015, available in
  httpsgithub.commauricioanicheck.

\bibitem{aleti2020apr}
A.~Aleti and M.~Martinez, ``E-apr mapping the effectiveness of automated
  program repair,'' \emph{arXiv preprint arXiv2002.03968}, 2020.

\bibitem{munoz2018instance}
M.~A. Mu{~n}oz, L.~Villanova, D.~Baatar, and K.~Smith-Miles, ``Instance spaces
  for machine learning classification,'' \emph{Machine Learning}, vol. 107,
  no.~1, pp. 109--147, 2018.

\bibitem{smith2014towards}
K.~Smith-Miles, D.~Baatar, B.~Wreford, and R.~Lewis, ``Towards objective
  measures of algorithm performance across instance space,'' \emph{Computers \&
  Operations Research}, vol.~45, pp. 12--24, 2014.

\bibitem{gholami2017support}
R.~Gholami and N.~Fakhari, ``Support vector machine principles, parameters, and
  applications,'' in \emph{Handbook of Neural Computation}.\hskip 1em plus
  0.5em minus 0.4em\relax Elsevier, 2017, pp. 515--535.

\bibitem{jin2019enre}
W.~Jin, Y.~Cai, R.~Kazman, Q.~Zheng, D.~Cui, and T.~Liu, ``Enre a tool
  framework for extensible entity relation extraction,'' in \emph{Proceedings
  of the 41st International Conference on Software Engineering Companion
  Proceedings}.\hskip 1em plus 0.5em minus 0.4em\relax IEEE Press, 2019, pp.
  67--70.

\bibitem{patel2009software}
C.~Patel, A.~Hamou-Lhadj, and J.~Rilling, ``Software clustering using dynamic
  analysis and static dependencies,'' in \emph{2009 13th European Conference on
  Software Maintenance and Reengineering}.\hskip 1em plus 0.5em minus
  0.4em\relax IEEE, 2009, pp. 27--36.

\bibitem{lutellier2015comparing}
T.~Lutellier, D.~Chollak, J.~Garcia, L.~Tan, D.~Rayside, N.~Medvidovic, and
  R.~Kroeger, ``Comparing software architecture recovery techniques using
  accurate dependencies,'' in \emph{2015 IEEEACM 37th IEEE International
  Conference on Software Engineering}, vol.~2.\hskip 1em plus 0.5em minus
  0.4em\relax IEEE, 2015, pp. 69--78.

\bibitem{tsantalis_chatzigeorgiou_2009}
N.~Tsantalis and A.~Chatzigeorgiou, ``Identification of move method refactoring
  opportunities,'' \emph{IEEE Transactions on Software Engineering}, vol.~35,
  no.~3, p. 347–367, 2009.

\bibitem{chong2015analyzing}
C.~Y. Chong and S.~P. Lee, ``Analyzing maintainability and reliability of
  object-oriented software using weighted complex network,'' \emph{Journal of
  Systems and Software}, vol. 110, pp. 28--53, 2015.

\bibitem{tzerpos1999mojo}
V.~Tzerpos and R.~C. Holt, ``Mojo a distance metric for software clusterings,''
  in \emph{Reverse Engineering, 1999. Proceedings. Sixth Working Conference
  on}.\hskip 1em plus 0.5em minus 0.4em\relax IEEE, 1999, pp. 187--193.

\bibitem{naseem_deris_maqbool_li_shahzad_shah_2017}
R.~Naseem, M.~B. Deris, O.~Maqbool, J.-P. Li, S.~Shahzad, and H.~Shah,
  ``Improved binary similarity measures for software modularization,''
  \emph{Frontiers of Information Technology \& Electronic Engineering},
  vol.~18, no.~8, p. 1082–1107, 2017.

\bibitem{chidamber1994}
S.~R. Chidamber and C.~F. Kemerer, ``A metrics suite for object oriented
  design,'' \emph{IEEE Transactions on Software Engineering}, vol.~20, no.~6,
  pp. 476--493, 1994.

\bibitem{scanniello2013class}
G.~Scanniello, C.~Gravino, A.~Marcus, and T.~Menzies, ``Class level fault
  prediction using software clustering,'' in \emph{2013 28th IEEEACM
  International Conference on Automated Software Engineering (ASE)}.\hskip 1em
  plus 0.5em minus 0.4em\relax IEEE, 2013, pp. 640--645.

\bibitem{scanniello2010using}
G.~Scanniello, A.~D'Amico, C.~D'Amico, and T.~D'Amico, ``Using the kleinberg
  algorithm and vector space model for software system clustering,'' in
  \emph{Program Comprehension (ICPC), 2010 IEEE 18th International Conference
  on}.\hskip 1em plus 0.5em minus 0.4em\relax IEEE, 2010, pp. 180--189.

\bibitem{tzerpos2000stability}
V.~Tzerpos and R.~C. Holt, ``On the stability of software clustering
  algorithms,'' in \emph{Program Comprehension, 2000. Proceedings. IWPC 2000.
  8th International Workshop on}.\hskip 1em plus 0.5em minus 0.4em\relax IEEE,
  2000, pp. 211--218.

\bibitem{runeson2012case}
P.~Runeson, M.~Host, A.~Rainer, and B.~Regnell, \emph{Case study research in
  software engineering Guidelines and examples}.\hskip 1em plus 0.5em minus
  0.4em\relax John Wiley \& Sons, 2012.

\bibitem{campbell2013sonarqube}
G.~A. Campbell and P.~P. Papapetrou, \emph{SonarQube in action}.\hskip 1em plus
  0.5em minus 0.4em\relax Manning Publications Co., 2013.

\bibitem{vanDeursen1999}
\BIBentryALTinterwordspacing
A.~van Deursen and T.~Kuipers, ``Identifying objects using cluster and concept
  analysis,'' in \emph{Proceedings of the 21st International Conference on
  Software Engineering}, ser. ICSE '99.\hskip 1em plus 0.5em minus 0.4em\relax
  New York, NY, USA: ACM, 1999, pp. 246--255. [Online]. Available:
  \url{httpdoi.acm.org10.1145302405.302629}
\BIBentrySTDinterwordspacing

\bibitem{Fokaefs}
\BIBentryALTinterwordspacing
M.~Fokaefs, N.~Tsantalis, E.~Stroulia, and A.~Chatzigeorgiou, ``Identification
  and application of extract class refactorings in object-oriented systems,''
  \emph{Journal of Systems and Software}, vol.~85, no.~10, pp. 2241--2260, Oct.
  2012. [Online]. Available: \url{httpdx.doi.org10.1016j.jss.2012.04.013}
\BIBentrySTDinterwordspacing

\bibitem{anquetil1999experiments}
N.~Anquetil and T.~C. Lethbridge, ``Experiments with clustering as a software
  remodularization method,'' in \emph{Reverse Engineering, 1999. Proceedings.
  Sixth Working Conference on}.\hskip 1em plus 0.5em minus 0.4em\relax IEEE,
  1999, pp. 235--255.

\bibitem{cui2011applying}
J.~F. Cui and H.~S. Chae, ``Applying agglomerative hierarchical clustering
  algorithms to component identification for legacy systems,''
  \emph{Information and Software Technology}, vol.~53, no.~6, pp. 601--614,
  2011.

\bibitem{shtern2009methods}
M.~Shtern and V.~Tzerpos, ``Methods for selecting and improving software
  clustering algorithms,'' in \emph{2009 IEEE 17th International Conference on
  Program Comprehension}.\hskip 1em plus 0.5em minus 0.4em\relax IEEE, 2009,
  pp. 248--252.

\end{thebibliography}
\end{document}